\documentclass[aps,prx,reprint,twocolumn,superscriptaddress,floatfix,nofootinbib,longbibliography,nobibnotes,noeprint]{revtex4-2}
\usepackage[sort&compress]{natbib}
\usepackage{graphicx}
\usepackage{amsmath,amsfonts,amssymb,amsthm}
\usepackage{physics}
\usepackage{epsfig}

\usepackage[makeroom]{cancel}
\usepackage[caption=false]{subfig} 
\usepackage{mathrsfs}
\usepackage[normalem]{ulem}
\usepackage{enumitem}

\usepackage[english]{babel}
\usepackage{dsfont}
\usepackage{float}

\usepackage{tikz}
\usepackage{nicefrac}
\usepackage{blindtext}
\usepackage{transparent}
\usepackage{comment}

\usepackage[dvipsnames]{xcolor}
\usepackage[
	bookmarks=true,
	colorlinks,
	linkcolor=MidnightBlue,
	urlcolor=RoyalBlue,
	citecolor=RoyalBlue,
]{hyperref}
\usepackage{orcidlink}

\newcommand{\s}[1]{\hat{\sigma}^{#1}}

\definecolor{mygold}{rgb}{0.5,0.6,0.7}

\definecolor{mypurple}{rgb}{0.6,0.25,0.7}

\newcommand{\orcid}[1]{${}^{\orcidlink{0009-0006-7012-3358}}$}

\newcommand{\ASC}{\affiliation{Department of Physics and Arnold Sommerfeld Center for Theoretical Physics (ASC), Ludwig Maximilian University of Munich, 80333 Munich, Germany}}
\newcommand{\MCQST}{\affiliation{Munich Center for Quantum Science and Technology (MCQST), 80799 Munich, Germany}}
\newcommand{\NOTT}{\affiliation{School of Physics and Astronomy, University of Nottingham, Nottingham, NG7 2RD, UK}\affiliation{Centre for the Mathematics and Theoretical Physics of Quantum Non-Equilibrium Systems, University of Nottingham, Nottingham, NG7 2RD, UK}}
\newcommand{\LEEDS}{\affiliation{School of Physics and Astronomy, University of Leeds, Leeds LS2 9JT, UK}}
\newcommand{\MPQ}{\affiliation{Max Planck Institute of Quantum Optics, 85748 Garching, Germany}}
\newcommand{\KHU}{\affiliation{Department of Physics, College of Science, Kyung Hee University, Seoul 02447, Republic of Korea}}
\newcommand{\equalcontrib}{\thanks{These authors contributed equally to this work.}}

\begin{document}
\title{Role of flavor degrees of freedom in quantum simulations of disorder-free localization}

\author{Yizhuo Tian\orcid{0009-0006-7012-3358}}
\equalcontrib
\ASC
\MCQST

\author{Jared Jeyaretnam\orcid{0000-0002-8316-9025}}
\equalcontrib
\LEEDS
\NOTT

\author{Tanmay Bhore\orcid{0000-0002-9304-7144}}
\equalcontrib
\LEEDS

\author{Zlatko Papi\'c\orcid{0000-0002-8451-2235}}
\email{z.papic@leeds.ac.uk}
\LEEDS

\author{Jad C.~Halimeh\orcid{0000-0002-0659-7990}}
\email{jad.halimeh@lmu.de}
\ASC
\MPQ
\MCQST
\KHU

\begin{abstract}
	A recent \texttt{Google Quantum AI} experiment [\href{https://www.science.org/doi/10.1126/science.adr9680}{Gyawali \textit{et al.}, Science \textbf{393}, 71 (2026)}] has exploited quantum parallelism to emulate disorder-averaged many-body dynamics, with conserved local degrees of freedom generating an effective disorder potential.
	We investigate how the local spectrum of these static variables controls localization in a flavor-extended ${\mathbb Z}_2$ lattice gauge theory, which maps onto a mixed-field Ising chain with $n$-level bond disorder.
	Combining finite-size spectral and entanglement diagnostics with infinite matrix-product state dynamics, we find a qualitative distinction between binary and multilevel disorder.
	For $n=2$, apparent localization ultimately gives way to thermalization; the long-lived transient arises from energy-scale separation, degenerate spectral towers, and approximate Hilbert-space fragmentation.
	By contrast, $n=4$ displays consistent localization signatures, including Poissonian level statistics, area-law eigenstate entanglement, nonthermal entanglement spectra, and persistent local memory over accessible times in the thermodynamic limit.
	Our results show that, despite its larger variance, binary disorder lacks the local amplitude diversity needed to suppress resonances.
	Thus, localization is governed not simply by disorder strength, but by the local disorder spectrum and the resulting resonant connectivity of the many-body Hilbert space.
\end{abstract}

\date{\today}
\maketitle

\section{Introduction}
\begin{figure*}[tbp]
	\centering
	\includegraphics[width=\textwidth]{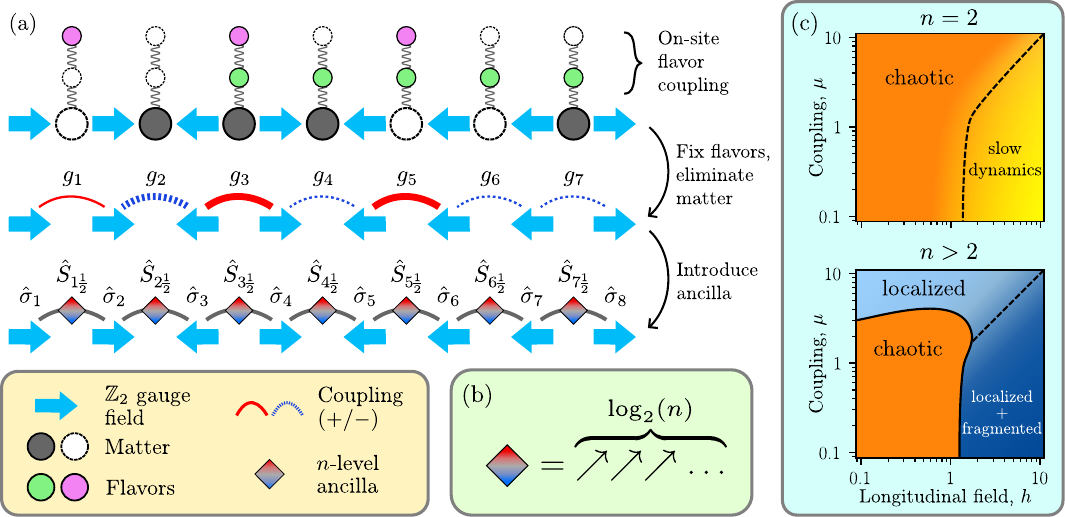}
	\caption{%
		(a)~Schematic of a $\mathbb{Z}_2$ lattice gauge theory and its mapping to an effective spin chain with $n$-level bond disorder.
		\textit{Top}:~$\mathbb{Z}_2$ gauge fields coupled to matter, with static external ``flavor'' particles coupled to the matter.
		\textit{Middle}:~After gauge fixing and eliminating matter fields (see Appendix~\ref{app:lgt-deriv}), the model maps to a mixed-field Ising chain with disordered couplings $g_j$, whose signs and magnitudes are set by the occupations of the flavor particles.
		\textit{Bottom}:~Introducing $n$-level ancilla $\hat{S}_{j+1/2}$ renders the model translationally invariant, enabling efficient disorder averaging (see Sec.~\ref{sec:iMPS}).
		(b)~Each $n$-level ancilla requires $\log_2(n)$ qubits, with consequences for experimental simulation of disordered models.
		(c)~Schematic summary of the phase diagrams obtained in this work as a function of longitudinal field $h$ and coupling $\mu$.
For $n=2$ the system remains chaotic throughout with slow dynamics at large $h$, while for $n>2$ localized and fragmented regimes emerge at large $\mu$ and $h$ respectively.
	}
	\label{fig:schematic}
\end{figure*}

A central question in nonequilibrium quantum physics is whether an isolated interacting many-body system approaches thermal equilibrium~\cite{PolkovnikovRMP,Eisert2015QuantumManyBodySystems}.
Generic nonintegrable systems are expected to thermalize according to the eigenstate thermalization hypothesis (ETH), losing local memory of their initial conditions at late times~\cite{Deutsch1991QuantumStatisticalMechanics,Srednicki1994ChaosAndQuantumThermalization,Rigol2008Thermalization,DAlessio2016FromQuantumChaos,Deutsch2018ETH}.
Thermalization can nevertheless be obstructed or strongly delayed by several mechanisms, including quantum many-body scars, kinetic constraints, and exact or approximate Hilbert-space fragmentation~\cite{Bernien2017,Turner2018WeakErgodicityBreaking,Serbyn2021QuantumManyBodyScars,Moudgalya2022QuantumManyBodyScarsHilbertSpaceFragmentation,Chandran2023QuantumManyBodyScars}.

A prominent route to avoiding thermalization is many-body localization (MBL), where sufficiently strong quenched disorder suppresses transport and produces quasi-local integrals of motion~\cite{Gornyi2005,Basko06,PalHuse,Serbyn2013LIOMs,Huse2014,Ros2015,Imbrie2014}.
Its characteristic signatures include persistent local memory, Poissonian level statistics, area-law entanglement in highly excited eigenstates, and slow entanglement growth following a quench~\cite{NandkishoreHuse,AbaninPapicAnnals,Abanin2019RMP,AletReview}.
Although the asymptotic stability of MBL remains under debate~\cite{DeRoeck2017StabilityDelocalizationMBL,Herviou2019MultiscaleEntanglementClusters,Gopalakrishnan2019InstabilityManybodyLocalized,Suntajs2020,AbaninChallenges,Crowley2022ConstructiveTheoryNumerically,Colmenarez2024ErgodicInclusions,Sels2023ThermalizationDiluteImpurities,Morningstar2022,Leonard2023,Ha2023ManyBodyLocalization,Biroli2024LargeDeviationAnalysis,Laflorencie2025CatStatesCarrying,Padhan2026LongRangeResonances} (see also the recent review~\cite{Sierant2025}), robust MBL signatures have been observed over experimentally and numerically accessible scales~\cite{Schreiber2015,Choi2016,Bordia2016,Smith2016,Roushan2017,Rispoli2019,LukinMBL2019,Yao2023,Stanley2023,Kjall2014,Luitz2015,BarLev2015,Mondaini2015,OBrien2016,Gray2018,Mierzejewski2018CountingLocalIntegrals,Theveniaut2020,Pietracaprina2021,Jeyaretnam2023}.

Distinguishing between genuine localization and finite-time nonergodic transient dynamics is challenging for classical simulations, which typically require averaging over many independently sampled disorder realizations.
Quantum parallelism offers an enticing alternative: disorder can be encoded in ancillary local degrees of freedom whose associated operators are conserved, hence their eigenvalues remain static during the evolution~\cite{Paredes2005ExploitingQuantumParallelism,Enss2017ManyBodyLocalizationInfiniteChains}.
Initializing the ancillas in an equal superposition of their eigenstates coherently encodes all disorder configurations, allowing the expectation values of physical observables to reproduce the disorder average within a \emph{single} unitary evolution.

Lattice gauge theories (LGTs) provide a natural realization of the above mechanism because their matter and gauge-field degrees of freedom are locally constrained by Gauss's law~
\cite{Wilson1974, Kogut_Susskind_LGT_PRD_1975, montvay1994quantum, Rothe2012LatticeGaugeTheories}.
The associated conserved local charges divide the Hilbert space into dynamically disconnected sectors;
within each sector, their fixed eigenvalues can act as an inhomogeneous background for the remaining dynamical degrees of freedom.
Consequently, a translation-invariant LGT Hamiltonian can display MBL-like dynamics without externally imposed disorder, a phenomenon known as disorder-free localization (DFL)~\cite{Smith2017AbsenceOfErgodicity, Smith2017DisorderFreeLocalization, Nandkishore2017, Metavitsiadis2017thermal, Akhtar2018, Papaefstathiou2020dfl, Brenes2018ManyBodyLocalization, Chanda2020, McClarty2020dfl, Hart2021logarithmic, Zhu2021subdiffusive, Karpov2021dfl, Sous2021PhononInducedDisorder, Halimeh2022enhancing, Chakraborty2022dfl, Gao2023, Osborne2023DisorderFreeLocalization, Hu2024, Cataldi2025DisorderFreeLocalizationFragmentation-1}.
The rapid development of quantum simulators for LGTs~\cite{Martinez2016RealtimeDynamicsLattice, Klco2018QuantumclassicalComputationSchwinger, Gorg2019RealizationDensitydependentPeierls, Schweizer2019FloquetApproachZ2, Mil2020ScalableRealizationLocal, Yang2020ObservationGaugeInvariance, Wang2022ObservationEmergent, Su2023ObservationManybodyScarring, Zhou2022ThermalizationDynamicsGauge, Wang2023InterrelatedThermalizationQuantum, Zhang2025ObservationMicroscopicConfinement, Zhu2024ProbingFalseVacuum, Ciavarella2021TrailheadQuantumSimulation, Ciavarella2022PreparationSU3Lattice, Ciavarella2023QuantumSimulationLattice-1, Ciavarella2024QuantumSimulationSU3, Gustafson2024PrimitiveQuantumGates, Gustafson2024PrimitiveQuantumGates-1, Lamm2024BlockEncodingsDiscrete, Farrell2023PreparationsQuantumSimulations-1, Farrell2023PreparationsQuantumSimulations, Farrell2024ScalableCircuitsPreparing, Farrell2024QuantumSimulationsHadron, Li2024SequencyHierarchyTruncation, Zemlevskiy2025ScalableQuantumSimulations, Lewis2019QubitModelU1, Atas2021SU2HadronsQuantum, ARahman2022SelfmitigatingTrotterCircuits, Atas2023SimulatingOnedimensionalQuantum, Mendicelli2023RealTimeEvolution, Kavaki2024SquarePlaquettesTriamond, Than2024PhaseDiagramQuantum, Angelides2025FirstorderPhaseTransition, Gyawali2025ObservationDisorderfreeLocalization, Mildenberger2025Confinement, Schuhmacher2025ObservationHadronScattering, Davoudi2025QuantumComputationHadron, Saner2025RealTimeObservationAharonovBohm, Xiang2025RealtimeScatteringFreezeout, Wang2025ObservationInelasticMeson, li2025frameworkquantumsimulationsenergyloss, mark2025observationballisticplasmamemory, froland2025simulatingfullygaugefixedsu2, Hudomal2025ErgodicityBreakingMeetsCriticality, hayata2026onsetthermalizationqdeformedsu2, Cochran2025VisualizingDynamicsCharges, Gonzalez-Cuadra2025ObservationStringBreaking, Crippa2024AnalysisConfinementString, De2024ObservationStringbreakingDynamics, Liu2024StringBreakingMechanism, Alexandrou2025RealizingStringBreaking, Cobos2025RealTimeDynamics2+1D, ilcic2026observationrobustcoherentnonabelian, chen2026thermalizationsu2latticegauge, Balaji:2025yua, Balaji:2025afl, xu2026observationglueballexcitationsstring, joshi2026observationgenuine21dstring, froland2026measuringnonstabilizernesssu2lattice}
has recently made such intrinsically generated disorder and its consequences for many-body dynamics experimentally accessible~\cite{Bauer2023,DiMeglio2024QuantumComputingHighEnergy,Halimeh2023ColdatomQuantumSimulators,Halimeh2025QuantumSimulationOutofequilibrium}.

In this work, we consider a $\mathbb Z_2$ LGT augmented by static gauge-neutral flavor degrees of freedom, as illustrated in Fig.~\ref{fig:schematic}(a).
After resolving the gauge constraints, the model maps onto a mixed-field Ising chain whose bond couplings are determined by the conserved flavor eigenvalues.
Each flavor configuration therefore defines an effective disorder realization, while a coherent superposition of configurations implements the disorder average.
The effective disorder is generated by the additional gauge-neutral flavor variables, similar to the recent \texttt{Google Quantum AI} realization of DFL through quantum parallelism~\cite{Gyawali2025ObservationDisorderfreeLocalization}.
Because experimentally encoded local variables have a finite-dimensional spectrum, however, the resulting disorder is necessarily discrete.
A two-level variable produces binary disorder, whereas an $n$-level variable generates $n$ distinct local couplings.
Realizing such a variable requires $\log_2 n$ ancillary qubits [Fig.~\ref{fig:schematic}(b)], creating a direct trade-off between the diversity of the effective disorder and the resources required to encode it.
This raises the central question of our work: what local spectrum must the static variables possess to generate robust localization rather than a long-lived nonergodic transient?

Previous studies have found MBL-like behavior with binary disorder and demonstrated that ancillary spins can implement disorder averaging directly in the thermodynamic limit~\cite{Enss2017ManyBodyLocalizationInfiniteChains}.
At the same time, comparisons among discrete disorder distributions indicate that multilevel disorder can reproduce the phenomenology of continuous randomness more closely, whereas binary disorder displays strong finite-size and boundary-condition effects~\cite{Janarek2018DiscreteDisorderModels}.
It remains unclear whether the binary disorder relevant to recent DFL experiments produces robust localization and, more generally, how localization depends on the local spectrum of the conserved variables.

Here, we address these questions by comparing binary disorder, $n=2$, with the next qubit-compatible realization, $n=4$, as well as with $n=8$ and continuous disorder.
We combine level statistics and eigenstate-entanglement diagnostics with real-time infinite-matrix-product-state dynamics.
Our principal findings are summarized in Fig.~\ref{fig:schematic}(c).
For $n=2$, apparent localization at finite sizes and intermediate times ultimately gives way to thermalizing behavior over the accessible converged scales.
We attribute the anomalously slow dynamics to degenerate spectral towers and the resulting approximate fragmentation of Hilbert space.
By contrast, $n=4$ and $n=8$ exhibit mutually consistent MBL-like signatures at sufficiently strong coupling, including Poissonian level statistics, area-law eigenstate entanglement, nonthermal entanglement spectra, and persistent local memory over accessible thermodynamic-limit times.
Thus, static conserved variables do not generically guarantee localization: their local spectrum controls the resonant connectivity of the many-body Hilbert space and thereby determines whether the system localizes or only exhibits a long-lived nonergodic transient.

The remainder of this paper is organized as follows.
In Sec.~\ref{sec:model}, we introduce the flavor-extended $\mathbb Z_2$ LGT  and derive its mapping onto a mixed-field Ising chain with $n$-level bond disorder.
In Secs.~\ref{sec:localization}-\ref{sec:iMPS}, we characterize ergodicity breaking using spectral, eigenstate, and dynamical diagnostics and explain the anomalously slow dynamics generated by binary disorder.
Finally, in Sec.~\ref{sec:discussion}, we summarize our results and discuss their implications for DFL and quantum-simulation protocols based on static conserved degrees of freedom.
Appendices contain further technical details of the mapping between different models and additional numerical results.

\section{Model}\label{sec:model}

We consider a one-dimensional flavor-extended $\mathbb{Z}_2$ LGT with dynamical matter fields on sites and gauge fields on links, as illustrated in Fig.~\ref{fig:schematic}(a).
The model is motivated by recent quantum-simulation experiments on DFL, in which a $\mathbb{Z}_2$ gauge structure and quantum parallelism were used to emulate disorder-averaged dynamics without explicitly sampling individual disorder realizations.
In particular, the $n=2$ version of a closely related model, together with a two-dimensional extension, was recently realized by \texttt{Google Quantum AI}~\cite{Gyawali2025ObservationDisorderfreeLocalization}.
The Hamiltonian is,
\begin{equation}
    \begin{aligned}
    \hat{H}_{\mathbb{Z}_2} ={} 
    & J\sum_j \hat{Z}_j\hat{\sigma}^z_{j,j+1}\hat{Z}_{j+1}
    + h\sum_j\hat{\sigma}^x_{j,j+1}\\
    & +\mu\sum_j\hat{X}_j\hat{D}_j^{(n)} .
    \end{aligned}
    \label{eq:H_flav}
\end{equation}
Here, $\hat{X}_j$ and $\hat{Z}_j$ act on the matter degree of freedom at site $j$, while $\hat{\sigma}^{x,z}_{j,j+1}$ act on the gauge field on the link between sites $j$ and $j+1$.
The operator $\hat{D}_j^{(n)}$ describes a static, gauge-neutral $n$-level flavor degree of freedom coupled to matter.
For $n=2^k$, it can be encoded using $k=\log_2n$ flavor qubits:
\begin{equation}
    \hat{D}_j^{(n)}
    =
    \sum_{a=0}^{\log_2(n)-1}
    \frac{2^a}{n-1}
    \left(2\hat{P}_j^{(a)}-1\right),
    \label{eq:Dj_flavor}
\end{equation}
where $\hat{P}_j^{(a)}$ projects onto the occupied state of the $a$th flavor qubit at site $j$.
The spectrum of $\hat{D}_j^{(n)}$ is therefore,
\begin{equation}
    D_j^{(n)} \in \left\{
    -1, -1+\frac{2}{n-1}, \ldots, +1-\frac{2}{n-1}, +1
    \right\},
    \label{eq:Dj_positive_values}
\end{equation}
consisting of $n$ equally spaced values between $-1$ and $+1$.
For example,
\begin{itemize}[noitemsep]
    \item $D_j^{(2)}\in\{-1,+1\}$,
    \item $D_j^{(4)}\in
    \big\{{-}1,-\tfrac{1}{3},+\tfrac{1}{3},+1\big\}$,
    \item $D_j^{(8)}\in
    \big\{{-}1,-\tfrac{5}{7},-\tfrac{3}{7},-\tfrac{1}{7},
    +\tfrac{1}{7},+\tfrac{3}{7},+\tfrac{5}{7},+1\big\}$.
\end{itemize}
Thus, increasing $n$ increases the number of effective disorder values at the cost of the additional ancillary qubits shown in Fig.~\ref{fig:schematic}(b).

Because the flavor variables are gauge neutral, they do not modify the local $\mathbb{Z}_2$ Gauss law,
\begin{equation}
    \hat{G}_j = -\hat{\sigma}^x_{j-1,j} \hat{X}_j \hat{\sigma}^x_{j,j+1},
    \qquad
    \hat{G}_j\ket{\psi} = (-1)^{q_j}\ket{\psi},
    \label{eq:Gauss_law}
\end{equation}
where $q_j\in\{0,1\}$ specifies the static background charge.
In the absence of flavor particles, $\hat{D}_j^{(n)}=\mathbb I$, Eq.~\eqref{eq:H_flav} reduces to the usual $\mathbb{Z}_2$ LGT  with a uniform matter mass.

As shown in Appendix~\ref{app:lgt-deriv}, fixing the gauge sector in Eq.~\eqref{eq:Gauss_law} and eliminating the matter fields maps the model onto a spin-$1/2$ mixed-field Ising chain,
\begin{equation}
    \hat{H}_0 = \mu \sum_{j=1}^{N-1} (-1)^{q_j+1} \hat{D}_j^{(n)}
    \hat{\sigma}_j^x \hat{\sigma}_{j+1}^x
    + h\sum_{j=1}^{N} \hat{\sigma}_j^x
    + J\sum_{j=1}^{N} \hat{\sigma}_j^z .
    \label{eq:hamiltonian_Dj}
\end{equation}
Here and below, $\hat{\sigma}_j^{x,z}$ denote the spins of the resulting Ising chain and should not be confused with the link operators in Eq.~\eqref{eq:H_flav}.
Since $\comm{\hat{H}_0}{\hat{D}_j^{(n)}}=0$, the flavor eigenvalues remain fixed during the evolution.
For a given configuration, we define $g_j^{(n)}=(-1)^{q_j+1}D_j^{(n)}$, which reduces Eq.~\eqref{eq:hamiltonian_Dj} to the disordered mixed-field Ising Hamiltonian,
\begin{equation}
    \hat{H} \big[\{g_j^{(n)}\}\big] = \mu \sum_{j=1}^{N-1} g_j^{(n)}
    \hat{\sigma}_j^x \hat{\sigma}_{j+1}^x
    + h\sum_{j=1}^{N} \hat{\sigma}_j^x
    + J\sum_{j=1}^{N} \hat{\sigma}_j^z .
    \label{eq:hamiltonian}
\end{equation}
Each configuration $\{g_j^{(n)}\}$ therefore defines one realization of $n$-level bond disorder.

To encode all disorder realizations simultaneously, we initialize the physical spins and flavor qubits in the product state,
\begin{equation}
    \begin{gathered}
        \ket{\Psi(0)} = \ket{\psi_0}_{\mathrm{sites}} \otimes \ket{\psi_F}_{\mathrm{flavor}},\\
        \ket{\psi_F}_{\mathrm{flavor}} = \bigotimes_{j,a} \frac{\ket{0}_{j,a} + \ket{1}_{j,a}}{\sqrt{2}}.
    \end{gathered}
    \label{eq:initial_state}
\end{equation}
The flavor state is an equal superposition of all eigenvalue configurations of $\{\hat{D}_j^{(n)}\}$.
Because different configurations do not mix under the evolution, expectation values of observables acting only on the physical spins reproduce the uniform disorder average generated by Eq.~\eqref{eq:hamiltonian}~\cite{Paredes2005ExploitingQuantumParallelism,Enss2017ManyBodyLocalizationInfiniteChains,Brenes2018ManyBodyLocalization,Jeyaretnam2025HilbertSpaceFragmentation,Gyawali2025ObservationDisorderfreeLocalization}.
In the limit $n\to\infty$, the discrete distribution approaches a uniform distribution on $[-1,1]$, commonly used in models of MBL.

We employ two complementary representations of the model in Eqs.~(\ref{eq:hamiltonian_Dj})-(\ref{eq:hamiltonian}).
For exact diagonalization studies in Secs.~\ref{sec:localization}-\ref{sec:fragmentation}, we sample fixed configurations $\{g_j^{(n)}\}$, evolve with Eq.~\eqref{eq:hamiltonian}, and explicitly average the results over disorder realizations.
For infinite-matrix-product-state simulations in Sec.~\ref{sec:iMPS}, we instead retain the flavor degrees of freedom as in Eq.~(\ref{eq:hamiltonian_Dj}) and evolve all configurations in parallel using the translationally invariant representation shown in the bottom panel of Fig.~1(a).
Finally, for $n=2$, Eq.~\eqref{eq:hamiltonian} can be mapped exactly onto a mixed-field Ising chain with binary \textit{site} disorder, as shown in Appendix~\ref{app:mapping}.

\begin{figure}[tbp]
    \centering
    \includegraphics[width=\linewidth]{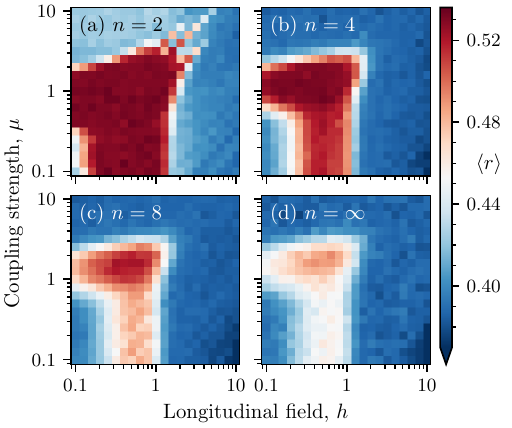}
    \caption{%
        Mean level-spacing ratio $\langle r\rangle$ as a function of the longitudinal field $h$ and effective disorder scale $\mu$, for the model in Eq.~(\ref{eq:hamiltonian}) at system size $N=16$ and different numbers $n$ of discrete disorder levels.
        Here, $n=\infty$ in panel (d) denotes a continuous uniform distribution $g\in[-1,1]$.
        Results are averaged over 300 disorder realizations and the 100 eigenstates closest to the peak of the density of states.
        The Poisson and GOE values are $\langle r\rangle_{\mathrm{P}}\approx0.386$ and $\langle r\rangle_{\mathrm{GOE}}\approx0.531$, respectively.
    }
    \label{fig:phase_diagram_r}
\end{figure}

\section{Spectral and entanglement signatures of localization}
\label{sec:localization}

We begin by comparing binary and multilevel disorder using two standard eigenstate diagnostics: level statistics and bipartite entanglement entropy.
Both are computed by exact diagonalization after resolving the relevant symmetries and targeting eigenstates near the peak of the density of states; further details are given in Appendix~\ref{app:ed}.

The level-spacing ratio is defined as,
\begin{equation}
    r_i = \min\left( \frac{s_i}{s_{i+1}}, \frac{s_{i+1}}{s_i} \right), \qquad
    s_i=E_{i+1}-E_i .
\end{equation}
After resolving all symmetries, its mean approaches $\langle r\rangle_{\mathrm{P}}\approx0.386$ for Poisson statistics, characteristic of integrable or localized systems, and $\langle r\rangle_{\mathrm{GOE}}\approx0.531$ for chaotic systems with time-reversal symmetry.

Figure~\ref{fig:phase_diagram_r} shows $\langle r\rangle$ across the $(h,\mu)$ plane, with $J=1$ and $0.1\leq h,\mu\leq10$.
We compare binary disorder, $n=2$, with $n=4$, $n=8$, and the continuous-disorder limit $n=\infty$.
At the finite size shown, all four cases contain broad regions in which $\langle r\rangle$ is suppressed towards the Poisson value.
Level statistics at a single system size could therefore suggest that binary disorder supports localization similar to that found for $n\geq4$.
However, as we show below, the scaling of eigenstate entanglement leads to a qualitatively different conclusion.

For $n=4,8,\infty$, $\langle r\rangle$ also remains somewhat below the GOE value in parts of the putative thermal regime.
We attribute these deviations primarily to finite-size effects associated with weak bonds.
For finite even $n$, the smallest coupling magnitude is $|g|_{\mathrm{min}}=1/(n-1)$, while for continuous disorder the typical weakest bond in a finite chain decreases algebraically with $N$.
Such bonds can behave as approximate cuts of the chain, suppressing level repulsion and entanglement at accessible sizes.
In the thermal phase, however, the many-body level spacing decreases exponentially with $N$, so a nonzero weak bond should eventually hybridize nearby many-body states.

\begin{figure*}[tbp]
    \centering
    \includegraphics[width=\linewidth]{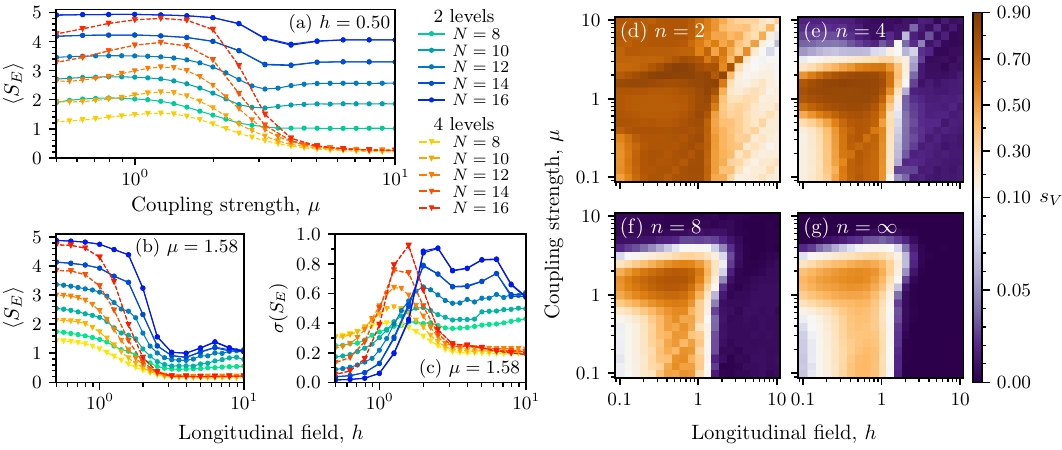}
    \caption{%
        Finite-size scaling of the bipartite entanglement entropy in many-body eigenstates of the model in Eq.~(\ref{eq:hamiltonian}).
        (a) Mean entanglement entropy $\langle S_E\rangle$ as a function of $\mu$ at $h=0.5$.
        Binary disorder ($n=2$) retains volume-law scaling, whereas $n=4$ crosses into an area-law regime at large $\mu$.
        (b) Corresponding results as a function of $h$ at $\mu=1.58$.
        Although the volume-law coefficient for $n=2$ becomes small at large $h$, its entropy continues to increase with $N$.
        (c) Standard deviation of $S_E$ within individual disorder realizations, followed by averaging over realizations.
        For $n=4$, the sharpening peak is consistent with a transition between volume- and area-law regimes.
        For $n=2$, the sharp peak is washed out by broad fluctuations throughout the large-$h$ regime.
        (d)-(g) Volume-law coefficient $s_V$ from Eq.~(\ref{eq:entropy_fit}) for $n=2,4,8,\infty$.
        Multilevel disorder develops regimes with $s_V\approx0$, whereas $n=2$ retains a nonzero volume-law contribution throughout the parameter range studied.
        Statistical uncertainties in panels (a--c) are smaller than the symbols.
    }
    \label{fig:S_scaling}
\end{figure*}

To distinguish localization from finite-size or finite-energy-scale effects, we examine the scaling of the bipartite entanglement entropy,
\begin{equation}
    S_E
    =
    -\Tr\left(\hat{\rho}_A\ln\hat{\rho}_A\right),
    \qquad
    \hat{\rho}_A
    =
    \Tr_B\ket{\psi}\!\bra{\psi},
\end{equation}
for an equal bipartition of the chain into subsystems $A$ and $B$.
Highly-excited thermalizing eigenstates  exhibit volume-law entanglement, approaching the Page value, $S_{\mathrm{Page}}=(N\ln 2-1)/2$, for a random state at large even $N$~\cite{Page1993AverageEntropySubsystem}.
By contrast, MBL eigenstates obey an area law~\cite{Bauer2013,Serbyn2013LIOMs,Huse2014}, so that $S_E$ remains finite as $N$ increases.

Figure~\ref{fig:S_scaling} shows results for $S_E$ of the 100 eigenstates closest to the peak of the density of states, averaged over disorder for $N=8-16$.
For $n=2$ and $N\leq12$, we average over all disorder configurations.
Otherwise, we use 1000 realizations for $N\leq14$ and 300 realizations for $N=16$.
Figure~\ref{fig:S_scaling}(a) shows $\langle S_E\rangle$ as a function of $\mu$ at fixed $h=0.5$.
The difference between binary and multilevel disorder is clear: for $n=2$, the entanglement continues to grow approximately linearly with $N$ throughout the range studied, while for $n=4$, the curves collapse towards an $N$-independent value at large $\mu$, consistent with area-law entanglement and localization.

The distinction is less immediate along the cut at fixed $\mu=1.58$ in Fig.~\ref{fig:S_scaling}(b).
The $n=4$ data again cross from volume-law to area-law scaling as $h$ increases.
For $n=2$, $\langle S_E\rangle$ is strongly suppressed at large $h$, which can mimic localization over small system sizes.
Nevertheless, it continues to increase with $N$, indicating a small but nonzero volume-law contribution rather than a stable area law.

Further information is provided by the fluctuations of $S_E$ within individual disorder realizations.
Figure~\ref{fig:S_scaling}(c) shows the corresponding standard deviation $\sigma(S_E)$, averaged over disorder.
For $n=4$, $\sigma(S_E)$ develops a peak that sharpens and grows with $N$, consistent with a transition into an area-law regime.
For $n=2$, the fluctuations increase around $h\approx2$ but remain broad up to the largest field studied.
Thus, even eigenstates close in energy retain a broad distribution of entanglement, with no clear convergence towards a stable localized regime.

We quantify the entropy scaling by fitting,
\begin{equation}
    \langle S_E(N)\rangle  = s_V \frac{N}{2} + c,
    \label{eq:entropy_fit}
\end{equation}
where $s_V$ is the volume-law coefficient.
Figures~\ref{fig:S_scaling}(d)--\ref{fig:S_scaling}(g) show $s_V$ across the $(h,\mu)$ plane.
For $n=4$, $n=8$, and $n=\infty$, extended regions with $s_V\approx0$ emerge at large $\mu$ or large $h$, consistent with area-law eigenstates.
These regions become particularly clear for $n=8$ and continuous disorder.
By contrast, for $n=2$ we find no comparable area-law regime: $s_V$ remains nonzero throughout the parameter range studied.
Its suppression at large $h$ explains why finite-size level statistics and entanglement can initially resemble localization, but the available scaling instead supports weak volume-law entanglement.

In summary, the level statistics and eigenstate entanglement reveal a qualitative distinction between binary and multilevel disorder.
Already for $n=4$, both diagnostics identify regimes consistent with MBL, whereas binary disorder produces strong finite-size signatures of nonergodicity without clear evidence of an area-law phase.
In the next section, we trace this anomalous binary behavior to degenerate spectral towers and approximate Hilbert-space fragmentation.

\section{Spectral towers and approximate Hilbert-space fragmentation}
\label{sec:fragmentation}

The preceding analysis shows that binary disorder produces strong finite-size signatures of nonergodicity without developing the area-law eigenstates found for $n\geq4$.
We now study in detail in the binary regime by focusing on the density of states (DOS) and the structure of individual eigenstates.
As we show below, binary disorder generates extensively degenerate spectral towers and an approximate decomposition of Hilbert space into weakly coupled manifolds.
This mechanism can strongly delay thermalization in finite systems while remaining distinct from MBL.

\subsection{Origin of the spectral towers}
\label{sec:towers}

\begin{figure}[tbp]
    \centering
    \includegraphics[width=\columnwidth]{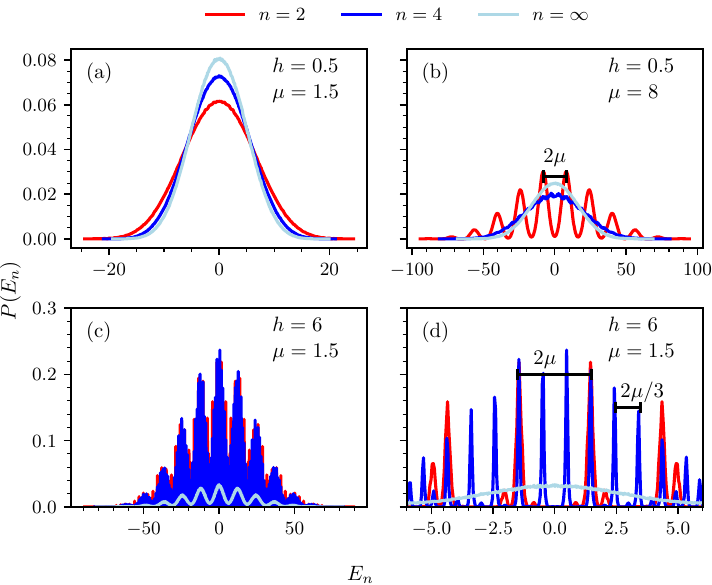}
    \caption{%
        Disorder-averaged density of states $P(E)$ for $N=12$ and the indicated values of $n$.
        (a) Thermal regime, $(h,\mu)=(0.5,1.5)$.
        (b) Strong-coupling regime, $(h,\mu)=(0.5,8)$.
        For finite $n$, the spectrum organizes into structures separated by multiples of $2\mu/(n-1)$; these are most pronounced for binary disorder.
        (c) Strong-field regime, $(h,\mu)=(6,1.5)$, where the principal towers are separated by $2h$.
        (d) Magnification of the central part of panel (c), revealing additional structure associated with the discrete bond energies.
        The $n=2$ results include all disorder configurations, while the other results are averaged over 1000 configurations.
    }
    \label{fig:DOS}
\end{figure}

Figure~\ref{fig:DOS} shows the disorder-averaged DOS at several representative points in the $(h,\mu)$ plane.
In the thermal regime, Fig.~\ref{fig:DOS}(a), the DOS has a smooth, approximately Gaussian form for all $n$.
At large $\mu$, however, the binary disorder spectrum separates into sharply resolved towers [Fig.~\ref{fig:DOS}(b)].
The towers become progressively less pronounced as the number of disorder levels increases, although residual structures remain visible for finite $n$.

The origin of the towers is most transparent in the limit $J=h=0$.
For a fixed disorder realization, the Hamiltonian reduces to,
\begin{equation}
    \hat{H}_{\mu}
    =
    \mu\sum_{j=1}^{N-1}
    g_j^{(n)}
    \hat{\sigma}_j^x\hat{\sigma}_{j+1}^x,
    \label{eq:H_strong_mu}
\end{equation}
whose eigenstates are product states in the $x$-basis.
For each bond, the lowest-energy spin configuration satisfies $g_j^{(n)}\sigma_j^x\sigma_{j+1}^x=-|g_j^{(n)}|$.
Relative to either of the two bond-satisfying ground states, a configuration containing defects on the set of bonds $\mathcal{D}$ has energy,
\begin{equation}
    E(\mathcal{D})
    =
    E_{\mathrm{GS}}
    +
    2\mu\sum_{j\in\mathcal{D}}|g_j^{(n)}|.
    \label{eq:defect_energy}
\end{equation}
Here, a defect denotes a bond whose spin configuration is reversed relative to its locally preferred value.

For binary disorder, $|g_j^{(2)}|=1$ on every bond, and the energy depends only on the number of defects $N_d=|\mathcal{D}|$:
\begin{equation}
    E(N_d)
    =
    E_{\mathrm{GS}}+2\mu N_d,
    \qquad
    D(N_d)
    =
    2\binom{N-1}{N_d}.
    \label{eq:binary_towers}
\end{equation}
The spectrum therefore consists of equally spaced, extensively degenerate towers whose degeneracy is maximal near $N_d=(N-1)/2$.
Importantly, randomizing the signs of the binary bonds does not lift this degeneracy because it leaves all bond magnitudes unchanged.
Indeed, binary bond disorder can be mapped exactly onto binary site disorder by a local unitary transformation, as shown in Appendix~\ref{app:mapping}.

For $n>2$, the defect cost $2\mu|g_j^{(n)}|$ depends on the bond on which the defect resides.
The degeneracy at fixed $N_d$ is therefore split according to the local bond magnitudes.
Because the allowed magnitudes remain commensurate for finite $n$, residual spectral structures separated by multiples of $2\mu/(n-1)$ survive.
For $n=4$, for example, the elementary defect energies are $2\mu/3$ and $2\mu$.
As $n$ increases, the number of distinct local energy scales grows and the tower structure becomes progressively less visible in the disorder-averaged DOS.

The continuous-disorder limit is qualitatively different.
The defect energies are continuously distributed, so exact degeneracies between prescribed configurations occur with probability zero and the disorder-averaged DOS becomes smooth.
The disappearance of spectral towers should not, however, be interpreted as a restoration of ergodicity.
Continuously distributed local energy mismatches can instead suppress resonant processes and support conventional disorder-induced localization.
Thus, the tower mechanism found for binary disorder and the localization produced by multilevel disorder have distinct microscopic origins.

Finite $h$ and $J$ broaden and hybridize the unperturbed towers.
The longitudinal-field term $h\sum_j\hat{\sigma}_j^x$ is diagonal in the same basis and splits states according to their $x$ magnetization, while $J\sum_j\hat{\sigma}_j^z$ generates transitions between $x$-basis configurations.
At large $\mu$, the latter predominantly hybridizes resonant configurations within a given tower or between nearby sublevels.
For binary disorder, moving a defect between bonds of equal magnitude does not change its leading-order energy, leaving large resonant manifolds.
The Hamiltonian consequently acquires an approximate block structure inherited from Eq.~\eqref{eq:binary_towers}, producing approximate Hilbert-space fragmentation and slow dynamics~\cite{Jeyaretnam2025HilbertSpaceFragmentation}.

A related hierarchy arises in the strong-field regime, $h\gg J,\mu$.
The leading Hamiltonian,
\begin{equation}
    \hat{H}_h
    =
    h\sum_j\hat{\sigma}_j^x
\end{equation}
produces magnetization towers separated by $2h$, as visible in Fig.~\ref{fig:DOS}(c).
Within each magnetization tower, the bond term resolves finer structures associated with the defect energies [Fig.~\ref{fig:DOS}(d)].
Because a single application of $\hat{\sigma}_j^z$ changes the $x$-magnetization, transitions within a fixed magnetization tower arise only through higher-order processes in $J/h$.
The spectrum is therefore approximately organized by both magnetization and defect energy.
This hierarchy explains why binary disorder exhibits particularly slow dynamics at large $h$.
It should not, however, be confused with exact strong fragmentation.
The tower manifolds remain coupled at finite $J/\mu$ or $J/h$, and the number of spectrally resolved manifolds is subexponential in the full Hilbert-space dimension.
The tower structure can therefore strongly delay thermalization but it does not  prevent it asymptotically.

The picture above also provides an interpretation of the narrow ``spike'' features visible in Figs.~\ref{fig:phase_diagram_r} and~\ref{fig:S_scaling} along lines of fixed ratios between $h$ and $\mu$.
These features arise when different tower spacings become commensurate, allowing higher-order processes to connect otherwise weakly coupled manifolds.
For example, near $h=\mu$, processes involving two neighboring spin flips can become resonant for an anomalously large number of configurations, enhancing hybridization and promoting delocalization.
The natural basis for this description is the $x$ basis.
In Appendix~\ref{app:ipr}, we study the participation ratio of midspectrum eigenstates in this basis.
For $n=4$, the eigenstates become strongly localized in the $x$-basis at both large $\mu$ and large $h$.
For $n=2$, by contrast, the participation ratio grows with system size, particularly at large $\mu$, demonstrating that the eigenstates remain extended over an increasing number of configurations despite the pronounced tower structure.

\subsection{Eigenstate structure within the towers}
\label{sec:tower_entanglement}

\begin{figure*}[tbp]
    \centering
    \includegraphics[width=\textwidth]{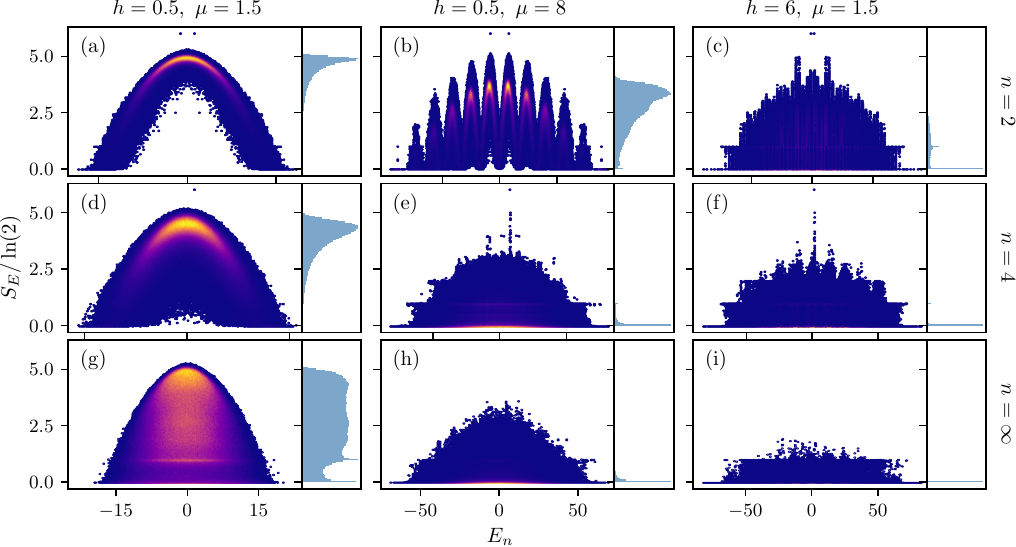}
    \caption{%
        Bipartite entanglement entropy $S_E/\ln 2$ of every eigenstate as a function of energy for $N=12$.
        Columns correspond to $n=2,4,\infty$, while rows show $(h,\mu)=(0.5,1.5)$, $(0.5,8)$, and $(6,1.5)$.
        Color denotes the local density of points, and the accompanying histograms show the distribution of $S_E$.
        Binary disorder retains broad distributions of entangled eigenstates in the strong-coupling and strong-field regimes.
        For multilevel and continuous disorder, most eigenstates instead accumulate near zero entanglement.
        All binary configurations and 1000 randomly sampled configurations for $n=4,\infty$ are included.
    }
    \label{fig:entanglement_stat}
\end{figure*}

The DOS identifies the tower structures in the energy spectrum but it does not tell us whether the eigenstates within each tower are thermal, localized, or fragmented.
Figure~\ref{fig:entanglement_stat} shows the entanglement entropy of every eigenstate as a function of its energy.
In the thermal regime, shown in the first row, midspectrum eigenstates concentrate near the Page value for all $n$.
At large $\mu$ or $h$, the distinction between binary and multilevel disorder becomes pronounced.

For $n=2$, the eigenstates remain broadly distributed over nonzero entanglement values within each spectral tower.
Their entropy is suppressed relative to the Page value because hybridization is largely restricted to a tower manifold, but it remains consistent with the volume-law scaling in Fig.~\ref{fig:S_scaling}.
Thus, the binary towers restrict the part of Hilbert space explored by the eigenstates without localizing them within that restricted manifold.
On the other hand, for $n=4$ and $n=\infty$, most eigenstates accumulate near $S_E=0$ in the strong-coupling and strong-field regimes.
This is consistent with the area-law scaling found in Fig.~\ref{fig:S_scaling}.
The comparison demonstrates that tower formation and localization are distinct phenomena: the most visible spectral towers occur for binary disorder, whereas the clearest area-law eigenstates occur for multilevel and continuous disorder.

In the continuous-disorder case, some eigenstates with $S_E\approx0$ or $S_E\approx\ln2$ also appear in the nominally thermal regime.
These states can be associated with exceptionally weak bonds that approximately divide the finite chain into two subsystems.
Nevertheless, the majority of midspectrum eigenstates remain close to the Page value, consistent with volume-law scaling in the thermal regime.

The mean entanglement entropy alone does not reveal how the entanglement is distributed among Schmidt values.
Two states with similar values of $S_E$ may therefore have qualitatively different internal entanglement structures.
To resolve this distinction, we consider the entanglement spectrum, i.e.\ the eigenvalues $\{\lambda_k\}$ of the reduced density matrix, ordered such that $\lambda_1\geq\lambda_2\geq\cdots\geq\lambda_d$, $\sum_{k=1}^{d}\lambda_k=1$.
Equivalently, $\lambda_k=s_k^2$, where $\{s_k\}$ are the Schmidt coefficients.
For a Haar-random pure state, the entanglement spectrum of an equal bipartition approaches the Marchenko--Pastur distribution~\cite{Yang2015TwoComponentStructure,Marcenko1967DistributionEigenvaluesSomeSets}.
Highly excited eigenstates of a chaotic Hamiltonian are therefore expected to approach this form with increasing system size.
By contrast, MBL eigenstates exhibit approximately power-law entanglement spectra as a consequence of their quasi-local integrals of motion~\cite{Serbyn2016PowerLawEntanglementSpectrum}.
The entanglement spectrum can therefore distinguish states that remain random-matrix-like within a restricted tower from states that are localized in the full Hilbert space.

To quantify deviations from random-matrix behavior, we compare the disorder- and eigenstate-averaged, rank-ordered entanglement spectrum $\{\lambda_k^{\mathrm{ES}}\}$ with a finite-dimensional, rank-ordered Marchenko--Pastur reference spectrum $\{\lambda_k^{\mathrm{MP}}\}$.
The latter is constructed for the same reduced Hilbert-space dimension and normalized such that $\sum_k \lambda_k^{\mathrm{MP}}=1$, see  Appendix~\ref{app:entanglement-spectrum}.
We calculate the Kullback--Leibler divergence,
\begin{equation}
    D_{\mathrm{KL}}
    \left(P_{\mathrm{MP}}\Vert P_{\mathrm{ES}}\right)
    =
    \sum_{k=1}^{d}
    \lambda_k^{\mathrm{MP}}
    \ln\left(
    \frac{\lambda_k^{\mathrm{MP}}}{\lambda_k^{\mathrm{ES}}}
    \right).
    \label{KL_div}
\end{equation}
A small $D_{\mathrm{KL}}$ indicates agreement with the random-matrix prediction, whereas a large value signals a nonthermal entanglement structure.

\begin{figure}[tbp]
    \centering
    \includegraphics[width=\columnwidth]{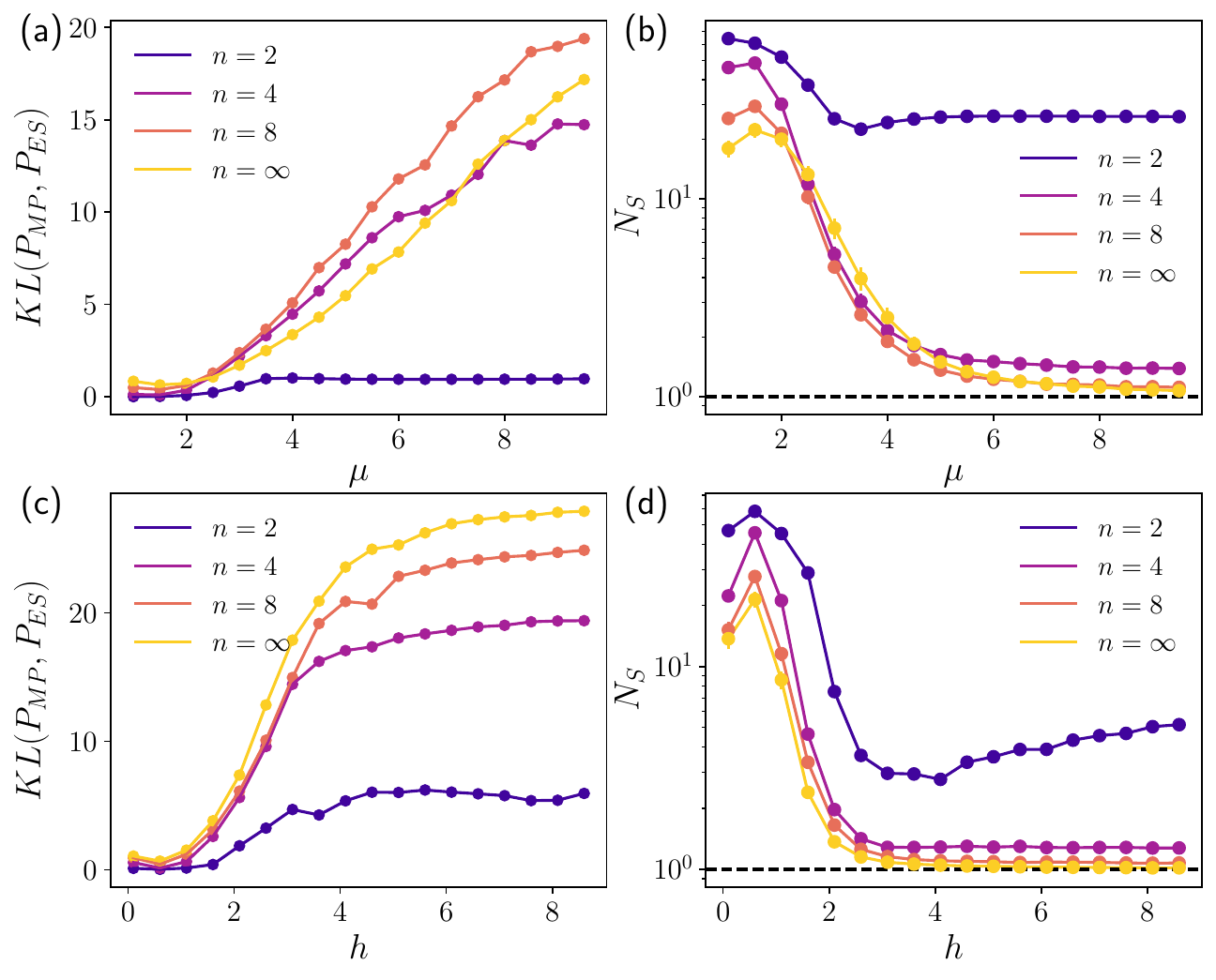}
    \caption{%
        Entanglement spectra for midspectrum eigenstates at $N=14$ and $J=1$.
        (a) KL divergence from the Marchenko--Pastur reference spectrum as a function of $\mu$ at fixed $h=0.5$.
        (b) Number $N_S$ of leading  eigenvalues required to reproduce $99\%$ of the entanglement entropy for the same parameters.
        (c)-(d) Corresponding results as a function of $h$ at fixed $\mu=1.58$.
        Results are averaged over the central $5\%$ of eigenstates for each of 100 disorder realizations.
    }
    \label{fig:entanglement_spectrum}
\end{figure}

Figures~\ref{fig:entanglement_spectrum}(a) and~(c) show $D_{\mathrm{KL}}$ as a function of $\mu$ and $h$, respectively.
At small $\mu$ and $h$, the entanglement spectra remain close to the Marchenko--Pastur prediction for all $n$, as expected in the thermal regime.
For $n=2$, the KL divergence remains comparatively small as either parameter is increased, although the departure from random-matrix behavior is more pronounced at large $h$ than at large $\mu$.
This is consistent with the stronger suppression of the binary eigenstate entanglement in the large-$h$ regime found in Fig.~\ref{fig:S_scaling}.

For $n=4,8,\infty$, by contrast, the KL divergence grows strongly with increasing $\mu$ or $h$.
The entanglement spectrum therefore departs from the random-matrix form precisely in the regimes where the mean eigenstate entanglement crosses towards area-law scaling.
Together, these observations show that the suppression of the mean entropy for $n=2$ and $n\geq4$ has different origins.
Binary eigenstates remain comparatively random-matrix-like within restricted tower manifolds, whereas multilevel disorder produces a strongly nonthermal entanglement structure.

We further characterize the entanglement spectrum by defining $N_S$ as the smallest number of leading eigenvalues $\lambda_k$ required to reproduce $99\%$ of the entanglement entropy,
$-\sum_{k=1}^{N_S}\lambda_k\ln\lambda_k/S_E
\geq0.99$.
For a thermal eigenstate, entanglement is distributed over an exponentially large number of Schmidt values and $N_S$ grows rapidly with system size.
For a strongly localized, nearly product eigenstate, only a small number of Schmidt values contribute appreciably and $N_S$ remains of order unity.

Figures~\ref{fig:entanglement_spectrum}(b) and~(d) show $N_S$ across the same parameter cuts.
At small $\mu$ and $h$, $N_S$ is large for all $n$, consistent with thermal eigenstates.
For $n=2$, it remains comparatively large as $\mu$ or $h$ is increased, again indicating that the eigenstates remain extended within the tower manifolds.
For $n=4,8,\infty$, $N_S$ instead decreases towards unity in the strong-coupling and strong-field regimes, consistent with eigenstates approaching localized product states in the $x$-basis.

In summary, the DOS, participation ratios, eigenstate entropies, and entanglement spectra further support two distinct mechanisms of slow dynamics in our model, Eq.~(\ref{eq:hamiltonian}).
For binary disorder, equal bond magnitudes produce degenerate spectral towers and approximate fragmentation, but the eigenstates remain extended.
For multilevel disorder, the diversity of local defect energies lifts these degeneracies and suppresses resonant hybridization, producing area-law eigenstates with nonthermal entanglement spectra.
The classical simulations of the dynamics directly in the thermodynamic-limit in the next section provide a direct test of whether these contrasting eigenstate structures lead to transient or persistent memory of the initial state.

\begin{figure*}[p]
	\centering
	\includegraphics[width=\linewidth]{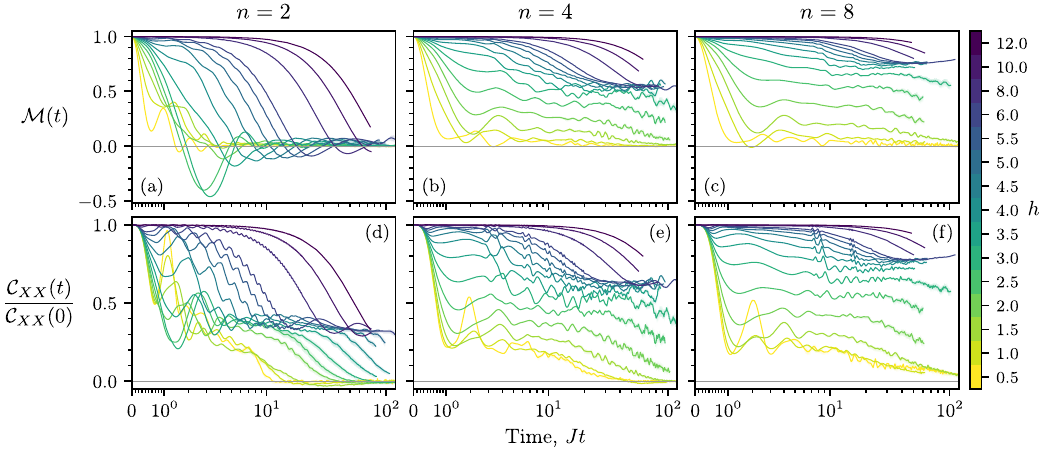}
	\caption{%
		Dynamics obtained by iTDVP for varying numbers of disorder levels $n$ and increasing $h$, while keeping $\mu = 1.5$, $J = 1.0$ fixed.
		Panels (a)-(c) show the staggered N\'eel magnetization $\mathcal{M}(t)$,  Eq.~(\ref{eq:stag_mag}), while panels (d)-(f) show the (normalized) nearest-neighbor correlation $\mathcal{C}_{XX}(t)$, Eq.~(\ref{eq:xx_measurement}).
		We collect data for $\chi$ from 20 up to 400 and extrapolate in $1/\chi$ to obtain an approximate $\chi \to \infty$ limit, shown by the solid lines (see Appendix~\ref{app:imps}).
		The shaded area around these lines represents the calculated uncertainty in the extraction (see main text).
	}
	\label{fig:iMPS_vary_h}
\end{figure*}
\begin{figure*}[p]
	\centering
	\includegraphics[width=\linewidth]{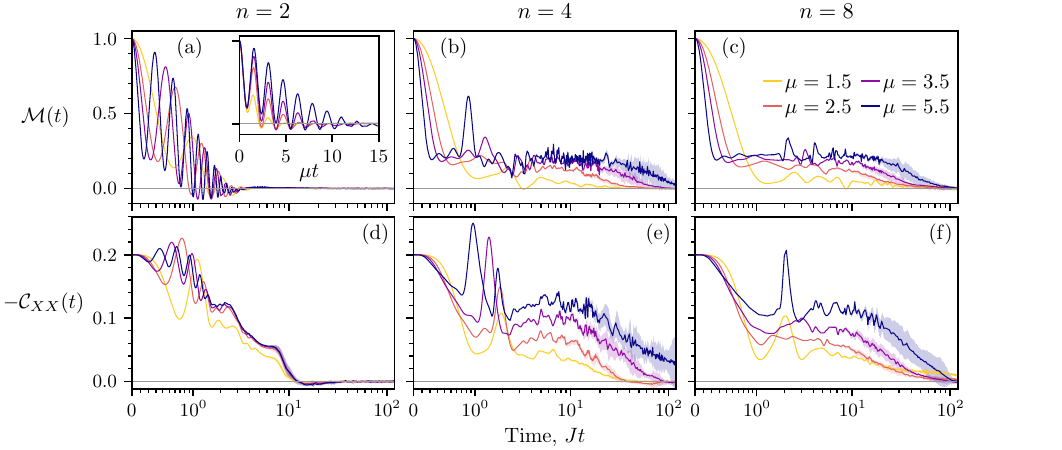}
	\caption{%
		Dynamics obtained by iTDVP for varying numbers of disorder levels $n$ and increasing $\mu$, while keeping $h = 0.5$, $J = 1.0$ fixed.
		Panels (a)-(c) show the staggered N\'eel magnetization $\mathcal{M}(t)$, Eq.~(\ref{eq:stag_mag}), while panels  (d)-(f) show the nearest-neighbor correlation $\mathcal{C}_{XX}(t)$, Eq.~(\ref{eq:xx_measurement}).
		The solid lines indicate the data for bond dimension $\chi = 400$, while the shaded area represents twice the difference between this data and that for $\chi = 300$, approximating the error induced by truncation;
		note that for $n \geq 4$ the errors become large after $Jt \simeq 10$.
	}
	\label{fig:iMPS_vary_mu}
\end{figure*}
\begin{figure}[tbp]
	\centering
	\includegraphics[width=\linewidth]{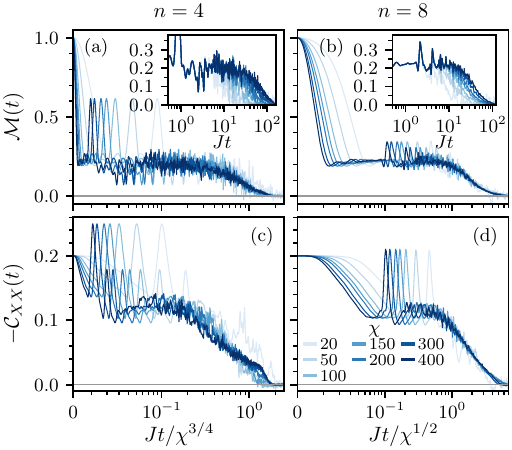}
	\caption{%
		iTDVP simulations at the point $\mu = 5.5$, $h=0.5$, $J=1.0$, for varying bond dimension $\chi$, for $n = 4, 8$.
		The early-time dynamics (before $Jt \simeq 10$) is converged in bond dimension, as shown in the insets, and we observe a clear plateau in magnetization $\mathcal{M}(t)$ and nearest-neighbor correlation $\mathcal{C}_{XX}$.
		However, at later times, the traces diverge, with magnetization and correlations decaying to zero, on a timescale increasing with $\chi$.
		We show in the main panels that by appropriately scaling the time with powers of $\chi$, this decay collapses to a single curve, suggesting that the plateau extends to late times in the limit $\chi \to \infty$.
		\label{fig:iMPS_vary_chi}
	}
\end{figure}

\section{Dynamics in the thermodynamic limit}
\label{sec:iMPS}

The eigenstate diagnostics above are necessarily restricted to finite systems.
We now investigate the dynamics directly in the thermodynamic limit using infinite matrix-product states (iMPS).
Rather than sampling fixed disorder configurations $\{g_j^{(n)}\}$ and evolving separately under Eq.~\eqref{eq:hamiltonian}, we retain the flavor degrees of freedom in Eq.~\eqref{eq:hamiltonian_Dj}.
Initializing them in an equal superposition of all flavor configurations then implements the uniform disorder average within a single translationally-invariant quantum evolution.
We simulate this enlarged system using the time-dependent variational principle for infinite MPS (iTDVP)~\cite{CiracRMP}.

We consider quenches from the initial product state $\ket{\Psi(0)}$ introduced in Eq.~\eqref{eq:initial_state}, where $\ket{\psi_F}_{\mathrm{flavor}}$ is the equal superposition of all flavor configurations.
The physical spins in $\ket{\psi_0}$ are alternately aligned and antialigned with the local field $h\hat{\sigma}^x+J\hat{\sigma}^z$:
\begin{equation}
    \hat{m}_j\ket{\psi_0} = (-1)^{j-1} \ket{\psi_0}, \qquad
    \hat{m}_j = \frac{h \hat{\sigma}_j^x+J \hat{\sigma}_j^z}{\sqrt{h^2+J^2}}.
    \label{eq:tilted_mag}
\end{equation}
The contributions of the local-field terms cancel within each two-site unit cell.
For a fixed disorder configuration, the remaining energy expectation value is,
\begin{equation}
    \bra{\psi_0} \hat{H} \big[\{g_j^{(n)}\}\big] \ket{\psi_0} = -\frac{\mu h^2}{h^2+J^2} \sum_{j=1}^{N-1}g_j^{(n)}.
    \label{eq:initial_energy}
\end{equation}
For a typical unbiased disorder realization, the sum on the right-hand side grows only as $\sqrt{N}$.
The initial energy density therefore vanishes as $N\to\infty$, placing the state near the center of the many-body spectrum.
In the ergodic regime, the state is consequently expected to relax rapidly towards the infinite-temperature ensemble.

Our iMPS representation uses a four-site unit cell containing two physical spin-$1/2$ degrees of freedom and two $n$-level flavor ancillas.
The operator $\hat{D}_j^{(n)}$ is proportional to the $S^z$ operator of a spin-$(n-1)/2$ degree of freedom, allowing the full disorder ensemble to be represented locally.
We assess convergence by repeating the simulations for bond dimensions up to $\chi=400$, see Appendix~\ref{app:imps}.

We monitor the tilted staggered magnetization,
\begin{equation}
    \mathcal{M}(t) = \frac{1}{2} \sum_{\ell=1,2} (-1)^{\ell-1} \langle \hat{m}_\ell(t) \rangle
    \label{eq:stag_mag}
\end{equation}
and the nearest-neighbor $x$-correlation,
\begin{equation}
    \mathcal{C}_{XX}(t) = \frac{1}{2} \sum_{\ell=1,2} \left\langle \hat{\sigma}_\ell^x \hat{\sigma}_{\ell+1}^x \right\rangle.
    \label{eq:xx_measurement}
\end{equation}
Expectation values in the enlarged physical-flavor state are equivalent to tracing out the flavor ancillas and averaging over disorder configurations.
Translation invariance then gives the corresponding disorder-averaged observables directly in the thermodynamic limit.
For the chosen initial state, $\mathcal{M}(0)=1$ and $\mathcal{C}_{XX}(0)=-h^2/(h^2+J^2)$.
Both observables have vanishing infinite-temperature expectation values, so persistent nonzero values at late times diagnose memory of the initial state.

We first consider the strong-field regime.
Figure~\ref{fig:iMPS_vary_h} shows the dynamics for $J=1$, $\mu=1.5$, and increasing $h$.
For these data, we extrapolate finite-$\chi$ results linearly in $1/\chi$, as described in Appendix~\ref{app:imps}.
At small $h$, both $\mathcal{M}(t)$ and $\mathcal{C}_{XX}(t)$ rapidly decay towards zero for all $n$, consistent with thermalization.

The behavior changes as $h$ is increased.
For binary disorder $n=2$, relaxation becomes progressively slower, but both observables continue to decay towards their thermal values over the accessible extrapolated time window.
We find no indication of a stable nonthermal plateau.
This is consistent with the previous eigenstate analysis: the strong field produces approximate fragmentation and a long prethermal timescale, but the eigenstates retain a volume-law contribution to their entanglement.

For $n=4$ and $n=8$, both observables instead develop nonzero plateaus for $h\gtrsim4$.
The plateaus become visible around $Jt\sim20$ and persist throughout the converged simulation window.
Increasing $h$ delays the approach to the plateau but has a comparatively weak effect on its height.
Although the accessible times do not establish asymptotic localization, the persistent memory is consistent with the area-law eigenstates and Poissonian level statistics found for multilevel disorder.
Finite-size exact-diagonalization dynamics presented in Appendix~\ref{app:ed-dynamics} show that the corresponding plateaus persist to substantially longer times.

The finite-bond-dimension behavior deserves particular attention.
In conventional MPS simulations, insufficient bond dimension often suppresses the growth of entanglement and could artificially enhance MBL signatures.
Here, we observe the opposite tendency: finite $\chi$ causes the late-time observables to decay more rapidly towards their infinite-temperature values.
A plausible explanation follows from the use of flavor ancillas to encode the disorder ensemble.
Representing the correlations between each flavor configuration and the corresponding physical state requires a bond dimension that grows during the evolution.
Truncation discards part of this configuration-dependent information and can thereby reduce the effective distinction among disorder realizations.
It can also generate truncation-induced correlations between flavor ancillas mediated by the physical spins.
Both effects can produce artificial late-time relaxation.

This issue is particularly important in the strong-coupling regime.
Figure~\ref{fig:iMPS_vary_mu} shows results at fixed $J=1$ and $h=0.5$ for increasing $\mu$, using the largest available bond dimension $\chi=400$.
For $n=2$, the staggered magnetization and nearest-neighbor correlation decay rapidly towards zero for every value of $\mu$ studied.
The relaxation timescale depends only weakly on $\mu$, although $\mathcal{M}(t)$ develops oscillations whose frequency increases linearly with $\mu$, as shown in the inset of Fig.~\ref{fig:iMPS_vary_mu}(a).
These oscillations are naturally associated with the spectral-tower spacing $2\mu$ derived in Sec.~\ref{sec:towers}.
Their decaying envelope confirms that the tower structure delays and structures the relaxation without producing localization.

For $n=4$ and $n=8$, both observables develop nonzero intermediate-time plateaus at large $\mu$.
These plateaus persist throughout the time interval in which the $\chi=300$ and $\chi=400$ results agree.
At later times, however, the different bond dimensions begin to diverge and the observables eventually decay towards zero.
The raw finite-$\chi$ data alone therefore cannot determine whether this late-time decay is physical or induced by MPS truncation.

To distinguish these possibilities, Fig.~\ref{fig:iMPS_vary_chi} compares the late-time dynamics over a broad range of bond dimensions at $\mu=5.5$ and $h=0.5$.
The early-time formation of the plateau is converged in $\chi$.
At later times, the onset of the decay shifts systematically to later times as the bond dimension increases.
Moreover, the curves collapse approximately when time is rescaled as $t/\chi^{3/4}$ for $n=4$, and $t/\chi^{1/2}$ for $n=8$.
No analogous scaling is observed for $n=2$, whose relaxation is already converged throughout the thermalization process.
The scaling collapse indicates that the late-time decay for $n=4$ and $n=8$ is controlled by the finite bond dimension: its characteristic timescale diverges as $\chi$ increases.
Extrapolating this empirical scaling suggests that the nonthermal plateau survives to progressively later times in the limit $\chi\to\infty$.
This indicates that the decay visible at finite $\chi$ is a truncation effect and places the physical relaxation time beyond the converged iMPS window.
This interpretation is also supported by exact-diagonalization dynamics at finite sizes, which retain nonzero memory out to $Jt>10^4$; see Appendix~\ref{app:ed-dynamics}.

The dynamics in the thermodynamic limit therefore reinforce the distinction found in the eigenstate diagnostics in Secs.~\ref{sec:localization}-\ref{sec:fragmentation}.
For binary disorder, the spectral towers and approximate fragmentation produce slow relaxation and persistent oscillations, but both local observables ultimately approach their thermal values over the accessible converged scales.
For $n=4$ and $n=8$, large $h$ or $\mu$ produces long-lived memory that persists throughout the accessible thermodynamic-limit time window and whose apparent finite-$\chi$ decay is pushed to later times as $\chi$ increases.
Together with the Poissonian level statistics, area-law eigenstate entanglement, and nonthermal entanglement spectra, these results provide mutually consistent evidence for an MBL-like regime generated by multilevel disorder.

\section{Conclusions and discussion}
\label{sec:discussion}

We have shown that the localization properties of the $n$-flavor $\mathbb{Z}_2$ LGT depend qualitatively on the local spectrum of the conserved flavor variables, which furnish an effective disorder potential.
For binary disorder $n=2$, finite-size and intermediate-time diagnostics suggest localization, but this interpretation is not supported by their scaling.
The eigenstates retain volume-law entanglement and remain extended in the $x$-basis, while local observables ultimately relax over the accessible converged thermodynamic-limit time window.
We instead attribute the anomalously slow dynamics to energy-scale separation, degenerate spectral towers, and approximate Hilbert-space fragmentation.
By contrast, $n=4$ and $n=8$ display mutually consistent MBL-like signatures, including Poissonian level statistics, area-law eigenstate entanglement, nonthermal entanglement spectra, localization in the $x$-basis, and persistent local memory over the accessible times.
We emphasize that these signatures are present even in the thermodynamic limit, $N\to\infty$, albeit up to finite times accessible to the numerical simulations.

Crucially, the distinct behavior between different numbers of flavors cannot be explained by the variance of the disorder.
For the uniformly weighted distributions considered here, $\sqrt{\langle g^2\rangle}_{n=2}=1$,
$\sqrt{\langle g^2\rangle}_{n=4}=\sqrt{5}/3$,
$\sqrt{\langle g^2\rangle}_{n=8}=\sqrt{3/7}$,
$\sqrt{\langle g^2\rangle}_{n=\infty}=1/\sqrt{3}$.
Therefore,
binary disorder has the largest root-mean-square coupling, yet exhibits the weakest evidence of localization.
The key difference is that $n=2$ randomizes only the signs of the bonds while leaving all magnitudes identical.
Consequently, every bond defect has the same elementary energy cost, preserving extensively degenerate spectral towers and large manifolds of resonantly connected configurations.
For $n\geq4$, several bond magnitudes generate several local excitation energies.
This lifts the binary degeneracies and suppresses resonant hybridization, even though the overall RMS coupling is smaller.
Increasing $n$ also introduces weaker bonds that may act as local bottlenecks, but the contrast with $n=2$ shows that weak links or disorder variance alone cannot account for the observed localization.
What matters is the full local disorder spectrum and the resonant connectivity that it induces in the many-body Hilbert space.

This conclusion is consistent with the exact mapping in Appendix~\ref{app:mapping}, which transforms binary bond disorder into binary longitudinal-field disorder without changing the spectrum.
Binary disorder is also known to produce anomalous finite-size and boundary-condition effects in random-field Heisenberg chains~\cite{Janarek2018DiscreteDisorderModels}.
In that setting, much of the difference between multilevel and continuous distributions can be reduced by matching their variances, although binary disorder remains exceptional.
Our results provide a sharper counterexample to a variance-based interpretation: increasing the disorder variance actually weakens the observed localization.
The effect of discretizing disorder is therefore model dependent and cannot be characterized by a single measure of disorder strength.

The strong-field regime further illustrates the distinction between fragmentation and localization.
For all $n$, a large $h$ organizes the spectrum into approximately decoupled magnetization manifolds.
For $n=2$, this structure produces very slow but ultimately decaying dynamics.
For $n=4$ and $n=8$, it coexists with multilevel disorder that further suppresses resonant coupling within and between the manifolds.
The resulting dynamics is even more strongly suppressed, see Appendix~\ref{app:ed-dynamics}.
Although we cannot determine its asymptotic form, this regime demonstrates how fragmentation-induced constraints and disorder-induced localization can reinforce one another.

We emphasize that slow dynamics observed in this work should be distinguished from weak thermalization reported in the clean mixed-field Ising chain~\cite{Banuls2011StrongWeakThermalization,Alishahiha2025ThermalizationKrylovBasis}.
There, slow relaxation was associated with a longitudinally polarized initial state having anomalously large overlap with a small number of low-lying eigenstates.
Our initial state, Eq.~\eqref{eq:tilted_mag}, lies near the center of the many-body spectrum.
Its slow dynamics instead originates from spectral towers and approximate fragmentation extending throughout the bulk of the spectrum.

Several open questions remain.
Our results do not establish the asymptotic stability of MBL for $n=4$ or $n=8$, since rare resonances or thermal avalanches may become important beyond the accessible sizes and times.
It would be useful to determine how localization depends on nonuniform flavor weights and on the precise arrangement of the discrete levels, thereby separating the effects of variance, weak bonds, and local spectral diversity more systematically.
Extensions to higher dimensions, non-Abelian gauge theories, dynamical flavor variables, and experimentally relevant noise or gauge-breaking perturbations are further directions worth exploring.

Finally, our results have direct implications for experimental studies of DFL using quantum parallelism.
Conserved local variables generate an ensemble of spatially inhomogeneous effective Hamiltonians, but do not by themselves guarantee localization: binary variables can produce long-lived nonthermal dynamics while remaining compatible with eventual thermalization.
Increasing the encoding from one to two ancillary qubits, corresponding to $n=2$ and $n=4$, therefore does more than refine the disorder resolution; it changes the local excitation spectrum and resonant connectivity.
Beyond disorder variance, the number, spacing, and weights of the flavor levels are thus important physical design parameters for future quantum simulations of DFL in LGTs.

\begin{acknowledgments}
\footnotesize
Y.T.\ and J.C.H.\ acknowledge funding by the Max Planck Society, the Deutsche Forschungsgemeinschaft (DFG, German Research Foundation) under Germany’s Excellence Strategy – EXC-2111 – 390814868, and the European Research Council (ERC) under the European Union’s Horizon Europe research and innovation program (Grant Agreement No.~101165667)—ERC Starting Grant QuSiGauge.
J.J.\ acknowledges support by the Leverhulme Trust Grant RPG-2024-112.
T.B.\ and Z.P.\ acknowledge support by the Leverhulme Trust Research Leadership Award RL-2019-015, and EPSRC Grants EP/Z533634/1, UKRI1337, UKRI3780.
Computational portions of this work were undertaken on the AIRE HPC system at the University of Leeds.
This work is part of the Quantum Computing for High-Energy Physics (QC4HEP) working group.
\normalsize
\end{acknowledgments}

\appendix

\section{Derivation from \texorpdfstring{$\mathbb{Z}_2$}{Z2} lattice gauge theory with disorder}\label{app:lgt-deriv}

We consider a one-dimensional flavor-extended $\mathbb{Z}_2$ LGT with dynamical matter as defined in Eq.~\eqref{eq:H_flav}.
In the $x$-basis, the Hamiltonian takes the form,
\begin{equation}
    \begin{aligned}
    \hat{H}_{\mathbb{Z}_2} = J & \sum_j \hat{Z}_j \hat{\sigma}^z_{j,j+1}\hat{Z}_{j+1}\\
    &+ h\sum_j \hat{\sigma}^x_{j,j+1} + \mu\sum_j \hat{X}_j \hat{D}_j\,.
    \end{aligned}
\end{equation}
This Hamiltonian respects a local $\mathbb{Z}_2$ gauge symmetry generated by,
\begin{equation}
    \hat{G}_j = -\,\hat{\sigma}^{x}_{j,j-1}\,\hat{X}_j\,\hat{\sigma}^{x}_{j,j+1},
\end{equation}
such that $[\hat{H}_{\mathbb{Z}_2}, \hat{G}_j] = 0$ for all $j$.
Physical states therefore satisfy the Gauss’s law constraint $\hat{G}_j|\psi\rangle = (-1)^{q_j}|\psi\rangle$, restricting the dynamics to the gauge-invariant subspace.
Here, $q_j \in \{0,1\}$ defines the external static charge located at matter site $j$.
We can then eliminate the matter degree of freedom using Gauss's law~\cite{Surace2021ScatteringMesonsQuantumSimulators}.
To do this, define a unitary transformation,
\begin{equation}
\begin{alignedat}{2}
    &\hat{U} \;=\;
    \prod_j
    \Big[
        P^+_{j+1, j-1}\mathbb{I}_j
        + P^-_{j+1, j-1}\,\hat{Z}_j
    \Big], \\[6pt]
    &P^\pm_{j+1, j-1} \;=\;
    \frac{1}{2}
    \Big(
        1 \pm
        \hat{\sigma}^{x}_{j,j+1}
        \hat{\sigma}^{x}_{j-1,j}
    \Big).
\end{alignedat}
\label{eq:Z2_LGT_U}
\end{equation}
This transformation flips the matter qubit if the adjacent gauge links are anti-aligned.
One can then show:
\begin{equation}
\begin{alignedat}{2}
    \hat{U}\,\hat{X}_j\,\hat{U}^\dagger
    &=
    \hat{X}_j\,\hat{\sigma}^x_{j,j+1}\,\hat{\sigma}^x_{j-1,j}, \\[6pt]
    \hat{U}\,\hat{Z}_{j}\,\hat{\sigma}^z_{j,j+1}\,\hat{Z}_{j+1}\,\hat{U}^\dagger
    &=
   \hat{\sigma}^z_{j,j+1},  \\[6pt]
    \hat{U}\hat{G   }_j\hat{U}^\dagger
        &\equiv \hat{G}'_j =
       -\hat{X_j}\,.
\end{alignedat}
\end{equation}
We may now choose the transformed gauge constraint $\hat{G}'_j = (-1)^{q_j} = \pm 1$, such that the transformed matter fields are pinned, $\hat{X_j} = (-1)^{q_j}$.
Finally, the transformed Hamiltonian reads,
\begin{equation}\label{eq:transformed_ham}
\begin{alignedat}{2}
    \hat{U}\,\hat{H}_{\mathbb{Z}_2}\,\hat{U}^\dagger
    &=\;
    J\sum_{j=1}^{N}\hat{\sigma}^z_{j,j+1}
    + h\sum_{j=1}^{N}\hat{\sigma}^x_{j,j+1}\,, \\[4pt]
    &\quad
    + \mu\sum_{j=2}^{N} (-1)^{q_j+1}\,\hat{D}_j\,
      \hat{\sigma}^x_{j-1,j}\hat{\sigma}^x_{j,j+1}\,.
\end{alignedat}
\end{equation}
Here, we define $(-1)^{q_j + 1} D^{(n)}_j \equiv g^{(n)}_j$, as introduced in the main text.
This expression reproduces Eq.~\eqref{eq:hamiltonian} after an index shift.

It is also noteworthy that this higher-level disorder set cannot be interpreted as a generalization of $\mathbb{Z}_N$ LGT or as a higher-level truncation of the gauge degrees of freedom in $U(1)$ LGT.
Performing the analogous unitary transformation in a $\mathbb{Z}_N$ LGT yields,
\begin{equation}\label{eq:ZN_LGT_Ham}
    \begin{aligned}
        \hat{U}\hat{H}_{\mathbb{Z}_{N}} \hat{U}^\dagger
    	=&\,\frac{J}{2}\sum_j\big(\hat{\sigma}_{j,j+1}+\text{h.c.}\big)\\
    	&+ \frac{h}{2}\sum_j\big(\hat{\tau}_{j,j+1}+\text{h.c.}\big)\\
    	&+\frac{\mu}{2}\sum_j\big(\omega^{q_j}\,\hat{\tau}^\dagger_{j-1,j}\hat{\tau}_{j,j+1}+\text{h.c.}\big).
    \end{aligned}
\end{equation}
where $\sigma$ and $\tau$ are the $\mathbb{Z}_N$ clock operators, with $\sigma$ having eigenvalues $e^{2\pi i k/N}$ and $\tau$ ($\tau^\dagger$) cyclically raising (lowering) the local state.
Likewise, integrating out the matter field in $U(1)$ LGT leads to a PXP-like model with a local spin-$s$ Hilbert space determined by the gauge truncation, while integrating out the gauge field produces a generalized XX chain.
None of these mappings generate the higher-level disorder structure considered here.

\section{Exact diagonalization}\label{app:ed}

Using exact diagonalization, we can explicitly construct the full Hamiltonian and access its eigenstates, for systems up to about $N = 16$.
In order to minimize finite-size effects and contamination from band-edge states, we specifically focus on the peak in the DOS, which is not always in the middle of the spectrum (see Sec.~\ref{sec:localization}).
For $N \leq 12$, we obtain the full spectrum using dense exact diagonalization and find the peak exactly, through binning the energies and picking the center of the bin with the most states.
For $N > 12$, we use shift-invert techniques to obtain select eigenstates closest to a target energy, and so do not have \textit{a priori} access to the energies.
We therefore approximate the peak in the DOS by using the spectrum when $J=0$, which is easy to calculate directly~\cite{Jeyaretnam2025HilbertSpaceFragmentation}, then choose the closest 100 states.

In the case of discrete disorder, particularly for $n = 2$, correlations between disorder realizations can introduce additional symmetries, such as discrete translational symmetry and parity symmetry~\cite{Janarek2018DiscreteDisorderModels}.
To address this, we impose open boundary conditions (OBC) to explicitly break translational symmetry.
We further resolve parity symmetry by block-diagonalizing the Hamiltonian into symmetry sectors for $N \leq 12$, while for $N > 12$, i.e.\ where we use shift-invert, we ignore disorder realizations with this property -- these form a vanishing proportion in the thermodynamic limit, and are already less than $1\%$ of sectors for $N = 14$ and $n = 2$.

We note that if we take all $g_j \to -g_j$, we obtain $-\hat{H}$ up to a local rotation $\prod_j \s{y}_j$, which will exhibit the same physics at infinite temperature.
Likewise, taking $g_j \to g_{N - j +1}$ gives an identical model up to a reflection about the center.
Carefully accounting for multiplicities, this allows us to reduce the number of disorder realizations needed to study the full ensemble for $n = 2$ and $N \leq 12$ by roughly a factor of 4.
(For $n = 4, 8, \infty$ or $N > 12$, we employ random sampling of realizations without replacement).

\section{Mapping from bond to site disorder}\label{app:mapping}
For the case $n = 2$, i.e.\ $g_i = \pm 1$, it is possible to map the bond disordered Hamiltonian~\eqref{eq:hamiltonian} to one with site disorder.
Let us write,
\begin{equation}
    \hat{H} = \sum_{j=1}^{N-1} \mu_j \s{x}_j \s{x}_{j+1} + \sum_{j=1}^N h_j \s{x}_j + J \sum_{j=1}^N \s{z}_j\,,
\end{equation}
with $\mu_j = \pm 1$ and $h_j = 1$ (initially).
Then, the unitary transformation $\hat{\Sigma}^z_k = \prod_{j \leq k} \s{z}_j$ will take $\mu_k \to -\mu_k$, and $h_j \to -h_j$ for $j \leq k$, while leaving all other coefficients and the spectrum unchanged.
If we apply $\hat{\Sigma}_k$ for each $k$ where $\mu_k = -1$, then we obtain a model with all $\mu_k = +1$, but disordered $h_j = \pm 1$.
Specifically, $h_{k+1}$ and $h_{k}$ will have different signs if $\mu_k = -1$, so we see that antiferromagnetic bonds map to domain walls in $h$.

For $n > 2$, it is not possible to eliminate bond disorder using this method, although we can transform to a model with all $\mu_j > 0$ but $h_j = \pm 1$.

\begin{figure}
    \centering
    \includegraphics[width=\linewidth]{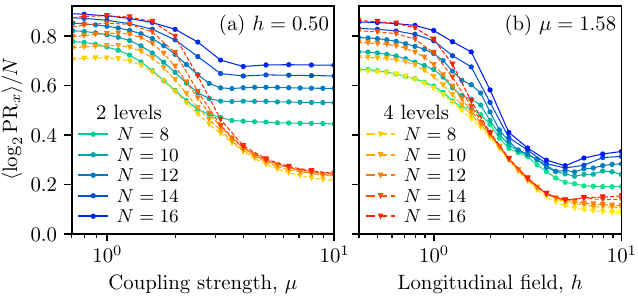}
    \caption{Participation ratio (PR) of mid-spectrum eigenstates in the $x$-basis, for (a) a cut through $\mu$ for $h = 0.50$, and (b) a cut through $h$ for $\mu$ = 1.58.
    We further take the logarithm (base 2) and scale by $1/N$, such that $0$ indicates complete localization and $1$ complete delocalization.
    Similarly to in Fig.~\ref{fig:S_scaling}, we see clear evidence of localization in the $x$-basis for $n=4$ for both large $h$, panel~(a), and large $\mu$, panel~(b).
    However, for $n=2$, the PR tends towards delocalization with increasing system size, especially clearly for large $\mu$ in panel~(a).
    [Error bars are smaller than the markers.]
    }
    \label{fig:IPR}
\end{figure}
\section{Localization in the \texorpdfstring{$x$}{x}-basis}\label{app:ipr}

We consider the participation ratio,
\begin{equation}
    \mathrm{PR}_x = \frac{1}{\sum_i^{2^N} |\!\braket{\psi}{x_i}\!|^4}\ ,
\end{equation}
where $\ket{x_i}$ is the $i$th $x$-basis state.
This roughly captures the number of states in the basis needed to faithfully represent $\ket{\psi}$: when it is $O(1)$, the state is fully localized in the basis, whereas when $\mathrm{PR}_x \sim O(e^N)$, the state is delocalized across exponentially many states.
Note that this is a basis dependent quantity.
In fact, $\mathrm{PR}_x = 1$ would imply complete delocalization in the $z$-basis.

In Fig.~\ref{fig:IPR}, we show the disorder-averaged logarithm of $\mathrm{PR}_x$, averaged over 100 eigenstates near the peak in the DOS.
We further normalize by $N$ such that 0 corresponds to complete localization ($\mathrm{PR}_x = 1$), and 1 to complete delocalization ($\mathrm{PR}_x = 2^N$) across the basis.
We see a dramatic difference between $n = 2$ and $n = 4$, especially for large $\mu$ in Fig.~\ref{fig:IPR}(a): while $\langle \log_2 \mathrm{PR}_x \rangle / N$ converges to a small value ${\sim}0.25$ for $n=4$, for $n=2$ we see that this value approaches $1$ for increasing $N$.
For large $h$ in Fig.~\ref{fig:IPR}(b), the difference is less dramatic: for both $n = 2, 4$, we see that $\langle \log_2 \mathrm{PR}_x \rangle / N$ becomes small.
But for $n=4$, the value gets much closer to $0$, and additionally for $n=2$ there is a substantial growth with $N$.

\section{Distribution of entanglement spectrum}
\label{app:entanglement-spectrum}

\begin{figure}[tbp]
    \includegraphics[width=\linewidth]{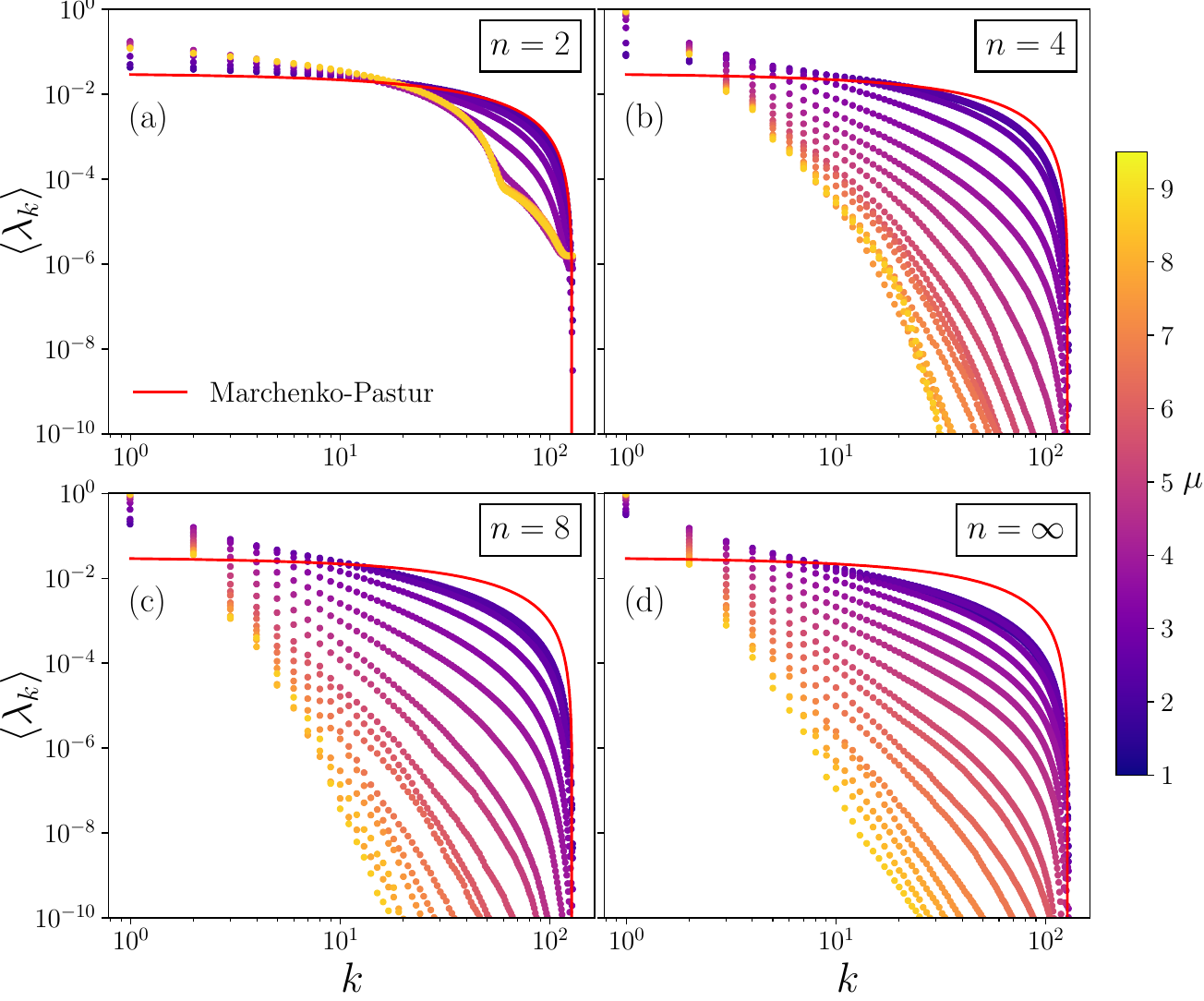}
    \caption{
        (a)-(d) Distribution of the entanglement spectrum of highly excited eigenstates as a function of $\mu$ for increasing number of levels $n$.
        The Marchenko--Pastur distribution is shown for reference.
        The data is for system size $N = 14$, $J = 1$ and $h = 0.5$ with a bipartition in the middle of the chain, averaged over $5\%$ of eigenstates and further averaged over $100$ charge sectors.
    }
    \label{fig:ES_appendix}
\end{figure}
The eigenspectrum of the reduced density matrix of a quantum state, known as its entanglement spectrum~\cite{LiHaldane}, can be found through a Schmidt decomposition state across a bipartition, $\ket{\Psi} = \sum_{k} s_k\ket{u_k}_A\otimes\ket{v_k}_B$.
The eigenvalues of the reduced density matrix, known as the entanglement spectrum are given by the square of the Schmidt coefficients $\lambda_k = s_k^2$ such that $\sum_k \lambda_k=1$.

In the spirit of random matrix theory~\cite{mehta2004random}, the entanglement spectrum of a random pure state is described by the Marchenko--Pastur distribution~\cite{Marcenko1967DistributionEigenvaluesSomeSets}.
This distribution gives the asymptotic eigenvalue density of a Wishart matrix $Y=BB^{\dagger}$, where $B$ is an $N\times M$ random matrix with independent and identically distributed entries.
For an equal bipartition, corresponding to $N=M=d$, the normalized reduced density matrix can be written as $\hat \rho_A=BB^{\dagger}/\mathrm{Tr}(BB^{\dagger})$, and its eigenvalues have support $0\leq \lambda\leq \lambda_{\mathrm{max}}$, with $\lambda_{\mathrm{max}}=4/d$ in the large-$d$ limit.
Let the eigenvalues be arranged in decreasing order,
$\lambda_1\geq \lambda_2\geq\cdots\geq p_d$,
and introduce the rescaled variable,
\begin{equation}
    x=\sqrt{\frac{\lambda}{\lambda_{\mathrm{max}}}} =\frac{1}{2}\sqrt{d\lambda},
    \qquad x\in[0,1].
\end{equation}
The complementary cumulative distribution associated with the Marchenko--Pastur density is,
\begin{equation}
    \eta(x) = 1-\frac{2}{\pi}
    \left[ x\sqrt{1-x^2} + \arcsin(x) \right],
\end{equation}
where $\eta(x)$ gives the fraction of eigenvalues larger than the value corresponding to $x$.

To construct the finite-dimensional, rank-ordered reference spectrum used in our numerical analysis, we introduce the uniformly spaced rank variables $\eta_k=\frac{k}{d}$, $k=1,\ldots,d$.
For every $\eta_k$, we numerically solve the equation $\eta(x_k)=\eta_k$ for $x_k\in[0,1]$ using a one-dimensional root-finding procedure.
The corresponding Marchenko--Pastur eigenvalues are then obtained as $\lambda_k^{\mathrm{MP}}=\lambda_{\mathrm{max}}x_k^2=(4/d)x_k^2$.
Since the prefactor $4/d$ ensures exact normalization only in the continuum limit, we remove the small finite-$d$ discretization error by normalizing the resulting spectrum with $\sum_{j=1}^{d}\lambda_j^{\mathrm{MP}}$.
The resulting list $\{\lambda_k^{\mathrm{MP}}\}$ is arranged in decreasing order and forms the normalized reference spectrum used throughout the analysis.
In the main text, we compare this spectrum with the disorder- and eigenstate-averaged reduced-density-matrix spectrum $\{\lambda_k^{\mathrm{ES}}\}$ using the KL divergence, Eq.~(\ref{KL_div}).

We show the full entanglement spectrum as a function of the eigenvalue rank in Fig.~\ref{fig:ES_appendix}.
The spectrum is averaged over a fraction of midspectrum eigenstates and over charge sectors, and is shown for increasing numbers of levels $n$.
The rank-ordered Marchenko--Pastur spectrum $\{\lambda_k^{\mathrm{MP}}\}$, generated using the procedure described above, is shown in red as a reference.
For $n=2$, the numerical entanglement spectrum closely follows the Marchenko--Pastur prediction at small and intermediate values of $\mu$, indicating agreement with the random-matrix prediction.
By contrast, the spectra for $n=4,8,\infty$ exhibit clear deviations from the Marchenko--Pastur form.

\section{Infinite matrix product state simulations}\label{app:imps}
For our dynamics simulations we make use of the uniform MPS ansatz~\cite{Zauner-Stauber2018VariationalOptimizationAlgorithms, Vanderstraeten2019TangentspaceMethodsUniform},
\begin{equation}
\label{eq:uniform_MPS}
    \ket{\Psi(A)}=\sum_{\vec{s}} \mathbf{v}_L^\dag \Big(\dots A^{s_{n-1}} A^{s_{n}} A^{s_{n+1}} \dots\Big) \mathbf{v}_R\,,
\end{equation}
where $A^s$ is a $\chi \times \chi$ matrix for each $s = 1, 2, \dots d$, for local Hilbert space dimension $d$.
The parameter $\chi$ is known as the bond dimension; larger $\chi$ allows the variational state to sustain greater correlations and longer-range entanglement, at the expense of increasing the computational complexity.
The boundary vectors $\mathbf{v}_L$ and $\mathbf{v}_R$ can be ignored as we consider  an infinite system and using an initial product state.
We may reinterpret $A^s$ as a $\chi \times d \times \chi$ tensor, in which case we can represent this state using standard tensor network notation as~\cite{CiracRMP}
\begin{equation*}
    \ket{\Psi(A)} =  \dots
    \begin{tikzpicture}[baseline = (X.base),every node/.style={scale=0.75},scale=.55]
    \draw (0.5,1.5) -- (1,1.5);
    \draw (1,2) rectangle (2,1);
    \draw (1.5,1.5) node (X) {$A$};
    \draw (2,1.5) -- (3,1.5);
    \draw (3,2) rectangle (4,1);
    \draw (3.5,1.5) node {$A$};
    \draw (4,1.5) -- (5,1.5);
    \draw (5,2) rectangle (6,1);
    \draw (5.5,1.5) node {$A$};
    \draw (6,1.5) -- (7,1.5);
    \draw (7,2) rectangle (8,1);
    \draw (7.5,1.5) node {$A$};
    \draw (8,1.5) -- (9,1.5);
    \draw (9,2) rectangle (10,1);
    \draw (9.5,1.5) node {$A$};
    \draw (10,1.5) -- (10.5,1.5);
    \draw (1.5,1) -- (1.5,.5);
    \draw (1.5, 0.25) node {$s_{n-2}$};
    \draw (3.5,1) -- (3.5,.5);
    \draw (3.5, 0.25) node {$s_{n-1}$};
    \draw (5.5,1) -- (5.5,.5);
    \draw (5.5, 0.25) node {$s_{n}$};
    \draw (7.5,1) -- (7.5,.5);
    \draw (7.5, 0.25) node {$s_{n+1}$};
    \draw (9.5,1) -- (9.5,.5);
    \draw (9.5, 0.25) node {$s_{n+2}$};
    \end{tikzpicture} \dots\,,
\end{equation*}
This represents a manifestly translationally invariant state, with a 1-site unit cell, although it can be straightforwardly generalized to 4-site unit cell used in this work.
By further casting this MPS into left- and right-canonical forms, and keeping track of the transformation between them, we can easily calculate expectation values of local observables.

We may then make use of the time-dependent variational principle (TDVP) which, as originally developed, finds the optimal time evolution of a dynamical system restricted to a manifold.
More recently, it has been extended to MPS, and to uniform MPS in particular~\cite{Zauner-Stauber2018VariationalOptimizationAlgorithms, Vanderstraeten2019TangentspaceMethodsUniform}.
We utilize a one-site update scheme with timestep $\Delta t = 0.01$, with all sites in the unit cell updated in parallel, and additionally attempt to expand the bond dimension $\chi$ across each bond at each time step until saturation.
These are implemented in the package \texttt{ITensorInfiniteMPS.jl}~\cite{LevyITensorInfiniteMPSjl, Fishman2022ITensor}.

To check the convergence of our simulations, we repeat our calculations at increasing $\chi = 20$, 50, 100, 150, 200, 300, 400.
We find that these give the exact same results (up to numerical precision) at short times before beginning to diverge, with the results for smaller $\chi$ diverging earlier (see Fig.~\ref{fig:iMPS_vary_chi}).
We therefore conclude our results represent the exact dynamics up to the time where the data for our two largest values of $\chi$ diverge.
However, we can extend the range of our simulations by extrapolating our finite-$\chi$ data.
This is done by fitting our data at each timestep to a straight line against $1/\chi$, and finding the intercept at $1/\chi = 0$, which corresponds to the limit $ \chi \to \infty$.
We find that if we exclude $\chi = 20, 50$, this straight line fit works very well for the large-$h$ regime in Fig.~\ref{fig:iMPS_vary_h}.
However, for the large-$\mu$ regime in Figs.~\ref{fig:iMPS_vary_mu}~\&~\ref{fig:iMPS_vary_chi} we find that the data at late times is best fitted by rescaling the time with $\chi$, as explained in the main text.

We also tested the convergence of our data with decreasing timestep $\Delta t$, by running simulations at $\Delta t = 0.005, 0.001$ for select $h$ and $\mu$.
We found here that for all times where the data is well converged with $\chi$ at fixed timestep, the difference between data with varying $\Delta t$ at fixed $\chi$ remained small (relative to the data and calculated uncertainties).
While decreasing the timestep undoubtedly increases the accuracy of our data, it also requires much more computational time, decreasing the simulation times we can reach, and we believe our choice of $\Delta t = 0.01$ strikes the best compromise between accuracy and computational efficiency.
\begin{figure*}[tbp]
    \includegraphics[width=1\textwidth]{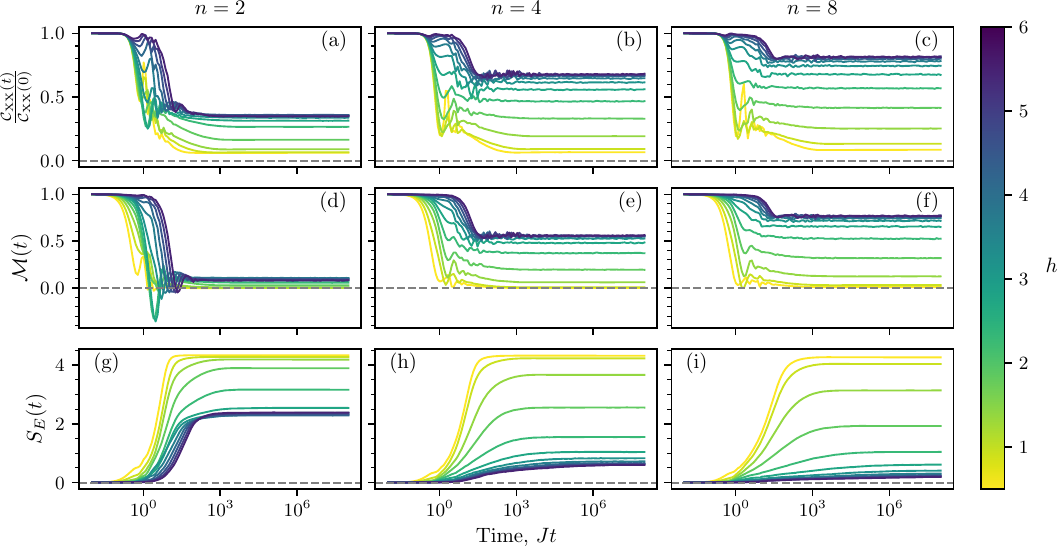}
    \caption{%
    	The long-time dynamics simulated via exact diagonalization with system size $N = 14$ for different disorder levels.
	    We compare the effect of varying the longitudinal field $h$ (with fixed $\mu = 1.5$, $J = 1.0$).
		Panels (a)-(c) show the nearest-neighbor normalized correlator $\mathcal{C}_{XX}(t)/\mathcal{C}_{XX}(0)$, Eq.~(\ref{eq:cxx_obc}),
        panels (d)-(f) show the staggered N\'eel magnetization $\mathcal{M}(t)$, Eq.~(~\ref{eq:stag_mag_obc}),   while panels (g)-(i) show the bipartite entanglement entropy $\mathcal{S}(t)$ after initializing in the tilted Néel state.
	    For $n = 2$, results are averaged over all disorder realizations, while for $n = 4, 8, \infty$ we randomly sample 1000 disorder realizations.
	   }
    \label{fig:EDdynamics_diffh}
\end{figure*}
\begin{figure*}[tbp]
    \includegraphics[width=1\textwidth]{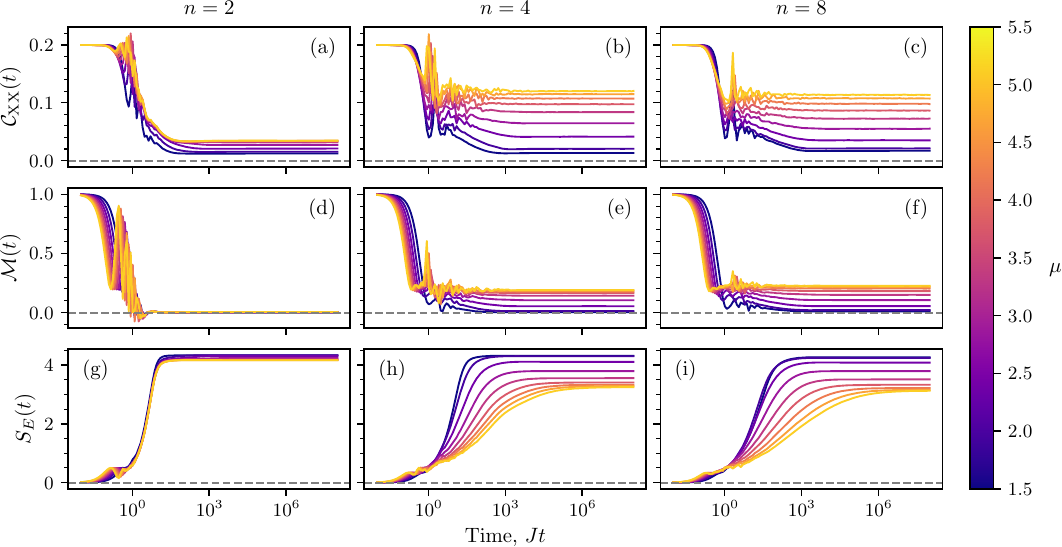}
    \caption{%
		Similar to Fig.~\ref{fig:EDdynamics_diffh}, but here we compare the effect of varying the coupling strength $\mu$ (with fixed $h = 0.5$, $J = 1.0$).
Data are obtained by exact diagonalization for system size $N=14$.
	}
    \label{fig:EDdynamics_diffmu}
\end{figure*}
\begin{figure}[tbp]
    \includegraphics[width=\linewidth]{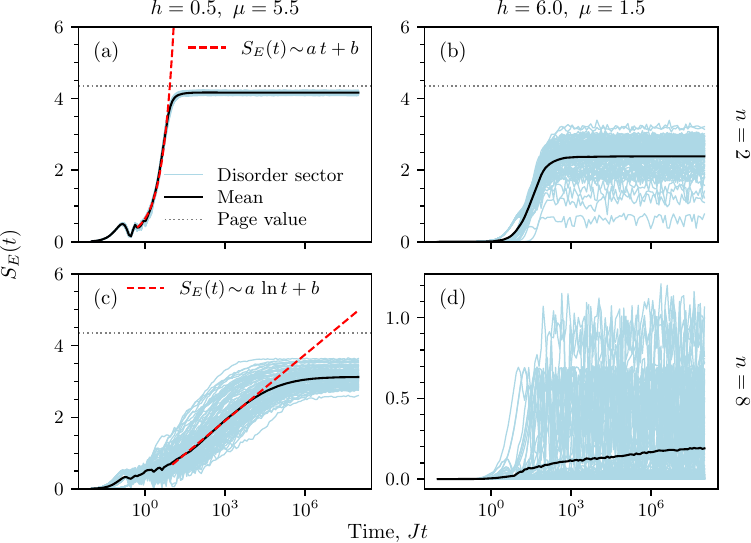}
    \caption{
        Comparison of the entanglement growth for $N = 14$ with $S_\text{Page}$ indicated by gray dotted lines, for $n=2$ and $n=8$
        in two distinct ergodicity-breaking regimes with $h = 0.5$, $\mu = 5.5$  and $h = 6.0$, $\mu = 1.5$.
        For each parameter set, 100 disorder sectors (light blue) and their mean (black) are shown.
        The red dashed line denotes the fitting function used to characterize the corresponding entanglement-growth behavior.
    }
    \label{fig:entropy_fit}
\end{figure}
\section{Long-time dynamics using exact diagonalization}\label{app:ed-dynamics}

To benchmark the \mbox{iTDVP} results and to investigate the long-time dynamics in the ergodicity-breaking regimes, we use exact diagonalization to perform quantum quenches from the tilted N\'eel state, Eq.~(\ref{eq:tilted_mag}), with open boundary conditions.
We use the same set of measurements as in the iMPS simulations for direct comparison,
\begin{gather}
    \mathcal{M}(t) =
    \frac{1}{N-2}\sum_{\ell = 2}^{N-1} (-1)^{\ell-1} \left\langle\Psi(t)\middle| \hat{m}_\ell \middle|\Psi(t)\right\rangle\,,
    \label{eq:stag_mag_obc}\\
    \mathcal{C}_{XX}(t) = \frac{1}{N-1}\sum_{\ell = 1}^{N-1}
    \langle \hat{\sigma}^x_\ell \hat{\sigma}^x_{\ell + 1} \rangle\,,
    \label{eq:cxx_obc}
\end{gather}
while excluding the boundary spins in the evaluation of $\mathcal{M}(t)$ in order to minimize boundary effects.

As shown in Fig.~\ref{fig:EDdynamics_diffh}, for the large-$h$ regime we observe a general tendency toward localization, indicated by the long-lived plateaus of both the nearest-neighbor correlation $\mathcal{C}_{XX}(t)$ in panels~(a)-(c) and the staggered magnetization $\mathcal{M}(t)$ in panels~(d)-(f).
We note that the case of $n=2$ exhibits a distinctly different behavior compared to higher-level disorder realizations, with the staggered magnetization reaching a steady state value close to zero.
This observation is mainly caused by the splitting of energy sectors and approximate fragmentation with finer internal structure, as discussed in Sec.~\ref{sec:fragmentation}.

In Fig.~\ref{fig:EDdynamics_diffmu}, we show another mechanism of ergodicity breaking emerging with increasing $\mu$.
For $n=2$, it is difficult to identify a clear localized regime upon increasing the coupling strength.
As seen in panel (d), the dynamics for larger $\mu$ remain qualitatively similar to those in the ETH regime, showing no persistent localization.
However, for $n=4, 8$, a pronounced localized plateau is maintained over accessible times, as shown in panels (b), (c), (e), and (f).
As illustrated in Fig.~\ref{fig:EDdynamics_diffmu}(h)-(i), the system exhibits a clear transition toward sub-ballistic entanglement-entropy growth with increasing~$\mu$, providing evidence for the onset of an MBL-like regime~\cite{Serbyn2013UniversalSlowGrowthEntanglement}.
We also observe that the ergodicity breaking induced by large~$\mu$ and~$h$ leads to qualitatively distinct entanglement behaviors.
Moreover, the dynamics are highly sensitive to whether the disorder level exceeds $n=2$, that is, whether effective disorder is actually present in the corresponding $\mathbb{Z}_2$ picture.

In Fig.~\ref{fig:entropy_fit}, we provide a detailed comparison of the entanglement-entropy dynamics across these different regimes.
For panels (a) and (b), corresponding to $n=2$, disorder does not play a significant role: the entanglement entropy grows linearly in time, consistent with rapid thermalization within each fragmented sector.
In the large $h$ regime, the entanglement for different disorder sectors saturates at distinct values below the Page value.
This occurs because, at large $h$, the initial state is close to the N\'eel state in the $\hat{\sigma}_x$ basis, and different disorder realizations project this state into fragmented sectors of varying Hilbert-space dimensions.
For large $\mu$ at $n = 4,8,\infty$, introducing multilevel disorder produces behavior consistent with an MBL-like regime over the accessible time scales, characterized by logarithmic entanglement growth with time.
In contrast, for large $h$, the dynamics are even more suppressed due to stronger fragmentation, resulting in slower than logarithmic growth, as shown in Fig.~\ref{fig:entropy_fit}(d).

\bibliography{biblio}
\end{document}